# The deleterious mutation load is insensitive to recent population history

Yuval B. Simons[1], Michael C. Turchin[2], Jonathan K. Pritchard[2,3,†] and Guy Sella[1,4,†]

May 4, 2013

[1]Department of Ecology, Evolution, and Behavior, The Hebrew University of Jerusalem

[2]Department of Human Genetics, [3]Howard Hughes Medical Institute, and [4]Department of Ecology and Evolution, The University of Chicago

[†]To whom correspondence should be addressed: pritch@uchicago.edu, gsella@math.huji.ac.il.

**Abstract**

Human populations have undergone dramatic changes in population size in the past 100,000 years, including a severe bottleneck of non-African populations and recent explosive population growth[1, 2, 3, 4, 5, 6, 7]. There is currently great interest in how these demographic events may have affected the burden of deleterious mutations in individuals and the allele frequency spectrum of disease mutations in populations[8, 4, 9, 7]. Here we use population genetic models to show that–contrary to previous conjectures–recent human demography has likely had very little impact on the average burden of deleterious mutations carried by individuals. This prediction is supported by exome sequence data showing that African American and European American individuals carry very similar burdens of damaging mutations. We next consider whether recent population growth has increased the importance of very rare mutations in complex traits. Our analysis predicts that for most classes of disease variants, rare alleles are unlikely to contribute a large fraction of the total genetic variance, and that the impact of recent growth is likely to be modest. However, for diseases that have a direct impact on fitness, strongly deleterious rare mutations likely do play important roles, and the impact of very rare mutations will be far greater as a result of recent growth. In summary, demographic history has dramatically impacted patterns of variation in different human populations, but these changes have likely had little impact on either genetic load or on the importance of rare variants for most complex traits.

Recent work has highlighted the impact of demographic history on the distribution of human genetic variation. Deep sequencing studies have identified huge numbers of very rare variants in human populations, the consequence of explosive population growth in the past five thousand years[2, 3, 7, 4, 5, 6]. Additionally, Europeans and east Asians have a greater fraction of high-frequency variants compared to Africans, likely due to an ancient bottleneck of non-African populations[10, 11, 1, 12, 6].

Given these observations, it is natural to ask whether recent demographic history has impacted the burden of genetic disease in modern human populations[8, 9, 7, 4]. Keinan and Clark[4] recently hypothesized that "Some degree of genetic risk for complex disease may be due to this recent rapid expansion of rare variants in the human population". A second important question concerns the relative importance of rare and common variants in causing disease[13, 14, 15]. If much of the genetic variation underlying disease is due to rare variants, then this could help to explain the so-called "missing heritability" of complex traits, and imply that mapping approaches based on deep sequencing will be essential for the dissection of complex traits[16].

To address these questions we analyzed a theoretical model with a large number of bi-allelic sites, each subject to two-way mutation, and natural selection against one of the alleles (see Supplement Section 1 for details). We studied three types of demographic models thought to be relevant for human populations: (i) a bottleneck; (ii) exponential growth starting from a constant-sized population; and (iii) a complex demographic model for African Americans (including rapid recent growth) and European Americans (including two bottlenecks followed by growth) inferred by Tennessen *et al.*[6]. The main features of the Tennessen model are similar to other recent models[17, 11, 12] while using a larger data set for parameter estimation. Our main results focus on selection against semi-dominant (i.e., additive) alleles in which the three genotypes have fitnesses 1, $1 - s/2$ and $1 - s$, respectively; and selection against recessive alleles with genotype fitnesses 1, 1, and $1 - s$. The effects of demography in these two models is qualitatively representative of those over the range of dominance coefficients (Supplement Section 2.4). In addition to simulation results shown here, further results and detailed theoretical analysis for all our key results are provided in the Supplement.

We focus first on the impact of demographic changes on individual load–that is, we want to understand whether demographic history has impacted the amount of deleterious variation carried by a typical individual in a population. Individual load is directly related to the number of deleterious alleles carried by an individual, or for recessive mutations to the number of homozygous sites per individual (see the Methods and Supplement for further details).

Figure 1 illustrates the impact of a bottleneck and population growth on the numbers of deleterious variants with strong selection ($s$=1%). As expected, these demographic events have a major impact on the number and frequency spectra of deleterious variants: the bottleneck causes a decrease in the total number of segregating sites in a population due largely to loss of rare variants, while the mean frequency of alleles that survive increases. Meanwhile, exponential growth causes a rapid increase in the number of segregating sites due to a major influx of rare variants, but a consequent drop in the mean frequency at segregating sites. But despite these dramatic shifts in the overall frequency spectrum, the impact on genetic load–namely, the mean number of deleterious variants per individual and thus the average fitness–is much more subtle.

In the semi-dominant case the load is essentially unaffected by these demographic events (Figures 1C and 1D). With growth, the increased number of segregating sites is exactly balanced by a decrease in mean frequency (and conversely for the bottleneck), so that the number of variants per individual stays constant. This kind of balance is predicted by classic mutation-selection balance models[18], and can be shown to hold for general changes in population size, provided that selection is strong and deleterious alleles are at least partially dominant (Supplement, Section 2.3).

The behavior of the recessive model is more complicated (Figures 1E and 1F). In the bottleneck model, the mean number of deleterious variants per individual drops by 60% as a result of the bottleneck. This is due to the loss of rare alleles. However, during the bottleneck, some deleterious alleles drift to higher frequencies[19, 8], contributing disproportionately to the number of homozygotes. This causes a transient increase in the number of deleterious homozygous sites per individual– i.e., load. Meanwhile, population growth has a less pronounced effect on recessive variation, leaving the mean number of deleterious alleles per individual unchanged, but causing a slight decrease in recessive load.

More generally, the manner in which demography affects load varies with the degree of dominance and the strength of selection (Figure 2 and Table 1, Supplement Section 2). The behavior of these models can be classified into three selection regimes (strong, weak and effectively neutral). In the strong selection case, i.e., where selection is much stronger than drift (approximately $s \geq 10^{-3}$ for semi-dominant), deleterious variants are extremely unlikely to fix, and virtually all of the genetic load is due to segregating variation. In this range, we infer that human demography has had no impact on semi-dominant load (and more generally for mutations with at least some dominance component), and will, if anything, have slightly reduced load due to recessive variation.

The weak selection case–where drift and selection have comparable effects–is more complex as fixed alleles may contribute appreciably to genetic load. This occurs over much longer time-scales than relevant here, however[20]. Specifically, the amount of fixation due to recent demographic changes is limited by both the time to fixation (on the order of $4N$ generations) and by the mutational input (on the order of $\frac{1}{2Nu}$ generations). For both semi-dominant and recessive variation, population growth is too recent to have substantially decreased the load due to fixed alleles. Recent growth increases the input of new mutations, but these mutations have had little time to move beyond very low frequencies and therefore have virtually no effect on the the total load. They do, however, turn sites that were previously fixed for the deleterious allele into segregating ones, increasing the contribution of segregating sites to load while reducing load due to fixed sites (Figure 2 and Supplement Section 2). The bottleneck in Europeans is estimated to have occurred further in the past and at much lower population sizes, allowing for some change in load at weakly selected sites. In this case, the increase in drift causes segregating deleterious alleles to increase in frequency, sometimes reaching fixation, resulting in a slight increase in load. The out-of-Africa bottleneck thus leads to a slight predicted increase of load in Europeans, most notably for recessive sites.

Finally, in the effectively neutral range–where selection has negligible effects on the population dynamics–segregating variation contributes negligibly to load and hence it does not change with demography. Thus, across all three selection regimes, recent human demographic history is likely to have had virtually no impact on genetic load at partially dominant sites, and only weak effects at recessive sites.

To evaluate these predictions using data, we considered single nucleotide variant (SNV) frequencies from a recent exome sequencing study of 2,217 African Americans (AAs) and 4,298 European Americans (EAs) sequenced at 15,336 protein coding genes (Fu *et al.*[7]; Figure 3). Previous work comparing numbers of deleterious variants in different populations has produced conflicting conclusions. For example, Lohmueller *et al.*[8] reported that there is "proportionally more deleterious variation in European than in African populations". Tennessen et al.[6] replicated this result using a conservative classification of deleterious sites, but found the opposite when using a more liberal classification of sites (both observations were highly significant).

To test whether there are differences in load between African and European Americans, we estimated the average numbers of derived alleles carried by individuals in each population at different functional classes of sites. Here, SNVs are defined as sites that are segregating in the entire sample of both AAs and EAs. Since we are using the derived allele count as a proxy for the deleterious allele count, there will be a low rate of misclassification at weakly selected sites that are fixed for the deleterious allele. However this does not change the qualitative predictions about patterns of differences between populations and we expect the number of derived alleles to have a monotonic relationship with the number of deleterious alleles. Specifically, for sites that are either neutral or semi-dominant, we predict that this measure should yield virtually identical counts in AAs and EAs (Figure 17, Supplement Section 3). At recessive sites, our model predicts slight differences (Supplement Section 3), but overall we expect these differences to be negligibly small. Note that when SNVs are defined within populations as in some previous papers, these simple predictions do not hold.

Functional predictions of SNVs were obtained from PolyPhen 2, a method that uses sequence conservation and structural information to infer which nonsynonymous changes are most likely to have functional consequences[21] (see Supplement Section 3 for similar analyses with other functional prediction methods). When using the functional predictions we observed a strong bias that most SNVs where the genome reference carries the derived allele are classified as benign, regardless of the overall population frequency (we were kindly alerted to this bias by D. Reich and S. Sunyaev, personal communication). Hence our analysis incorporates a correction to account for this effect (Supplement Section 3).

Figure 3 summarizes the results (see the Supplement for further analyses). As expected, the mean allele frequency declines with increasing functional severity[6], from 2.8% at noncoding SNVs to 0.6% at probably damaging SNVs, implying that there is selection against most SNVs with predicted damaging effects. More striking, however, is that within each of the five functional categories the mean allele frequencies–and hence the numbers of derived alleles per individual–are essentially identical in the two populations (p>.05 for all five comparisons). This observation is consistent with our model prediction that load should be very similar in these populations. Our result contradicts previously published results showing highly significant differences between AAs and EAs (e.g., Figure 4D of [6]); the discrepancy appears to be due to the reference bias in the functional predictions which, to our knowledge, has not been accounted for previously.

Although population size changes have had little impact on the average load carried by individuals, growth has greatly increased the number of rare variants. So do rare variants play a greater (and substantial) role in the genetics of disease as a result of recent growth (Figure 4)? Do higher frequency variants play a greater role in Europeans and Asians than in Africans? These questions are of practical importance because different mapping designs may be needed to identify rare variants[13, 16, 15, 22].

To study this, we computed the contributions of different allele frequencies to genetic variance (displayed here as the cumulative distribution of $x(1-x)f(x)/2$, where $f(x)$ is the probability that a derived allele is at frequency $x$). These distributions show the fraction of genetic variance for a disease that is contributed by alleles below frequency $x$, assuming that the loci underlying a trait are semi-dominant, with the indicated selection coefficient (Supplement Section 4). Note that in practice, we anticipate that variants underlying a given disease would have a variety of selection coefficients and effect sizes, in which case the overall distribution would be an appropriately weighted mixture of distributions for different selection coefficients.

Several interesting points are evident. For effectively neutral, or for weakly deleterious sites (Figure 4A), only a small fraction of the total variance comes from rare alleles: although there are many rare alleles, each one contributes very little to population variance and individual load. The same is true for recessive variation across almost the entire range of selection coefficients (Supplement Section 4.2). Likewise, if we assume that the frequency density $f(x)$ follows the frequency spectrum observed at all nonsynonymous sites classified as "probably damaging"[21] then, under the same model, it is still only a modest fraction of the genetic variance that is due to rare alleles (Figure 4B). Meanwhile, in all of these cases an Out-of-Africa bottleneck would increase the contribution of intermediate frequency alleles to the genetic variance (Figure 4A-C): e.g., at probably damaging sites 62% of the variance in EAs is contributed by alleles with minor allele frequency above 10% compared to only 49% in AAs.

It is only for the case of strong, dominant selection that rare variants become important (Figure 4C and 4D). For example, for a selection coefficient of 1%, most of the variation is rare and arose within the recent exponential growth phase. As a result, the contribution of extremely rare variants is much greater than it would have been in the absence of growth: e.g., in AAs and EAs, 80%, and 65% of the variance is due to alleles below frequency 0.1%, compared to just 25% in the constant population model.

Of course in practice, the genetic variants that contribute to a complex trait likely have a range of selection coefficients ($s$) and a range of effect sizes ($a$) on the phenotype in question (Supplement Section 4.3). When there is a mixture of selective coefficients, what can we say about the relative importance of rare and common variants? To answer this, the critical issue is to model the relationship between $a$ and $s$ [14, 23]. Here we consider two extreme models: (1) $a$ is independent of $s$ (this model would be most relevant for traits that are not closely tied to fitness, such as late-onset diseases or various quantitative traits); and (2) $a$ is proportional to $s$ (likely most relevant for traits with a direct impact on fitness such as early-onset diseases or diseases affecting fertility). Many traits presumably lie between these two extreme models.

Figure 4E shows the expected genetic variance per site as a function of $s$ under these two models. Under the constant-$a$ model, weakly selected mutations are much more likely to segregate at intermediate frequencies than strongly selected mutations, and hence these produce much higher variance–thus if sites that affect a disease have a range of different values of $s$, then we might expect the more weakly selected sites to contribute most of the genetic variance. But the reverse occurs in the model where $a$ increases with $s$: highly deleterious sites have the highest genetic variance due to their increased effect sizes.

To illustrate this further, Figure 4F shows the fraction of rare variants in a simple model in which the genetic variance for a disease is due to a mixture of weakly ($s = 0.0002$) and strongly ($s = 0.01$) selected mutations. As may be seen, under the constant-$a$ model rare variants make very little contribution to the overall genetic variance, unless nearly all disease variants are strongly selected. In contrast, when $a$ is proportional to $s$, most of the variance is due to rare alleles. Moreover, the contribution of rare alleles is much greater in this case than it would have been in the absence of recent population growth.

In summary, while recent demographic events have had well-documented effects on the frequency spectrum of SNVs in modern populations, we find that these events have had negligible impact on the average burden of mutations carried by individuals. Moreover, we conclude that although there are large absolute numbers of rare variants, they do not necessarily contribute a large fraction of the genetic variance underlying complex traits. It is only for diseases that are primarily due to strongly deleterious mutations that we can expect much of the variance to be due to rare alleles: these will likely tend to be diseases that are tightly coupled to fitness.

Literature cited.

# Acknowledgements

This work was supported by grants from the National Institutes of Health (MH084703, GM083228), the Israel Science Foundation (grant # 1492/10) and the Howard Hughes Medical Institute. Thanks to Molly Przeworski for comments and discussions, to David Reich and Shamil Sunyaev for helpful discussions and generous input regarding the interpretation of PolyPhen 2, and to Josh Akey for assistance in accessing data.

**Competing Interests.** The authors declare that they have no competing financial interests.

# Online Methods

This section provides a summary of our methods; a complete version may be found in the Supplementary Information.

**Model.** Our basic model considers selection at a single site. We use the standard bi-allelic diploid model with two-way mutation, viability selection, drift and, in some cases, migration [24]. Specifically, we assume there are two alleles at a site: normal ($N$) and deleterious ($D$). An $N$ allele mutates to the $D$ allele with probability $u$ per gamete, per generation and the reverse mutation occurs with probability $v$. Unless noted otherwise, we assume that mutation is symmetric, i.e., $u = v$. The absolute fitnesses of the three genotypes NN, ND and DD are 1, $1 - hs$ and $1 - s$, respectively, where $s > 0$ and $h \geq 0$.

We focus on semi-dominant ($h = \frac{1}{2}$) and fully recessive ($h = 0$) selection because these two cases exhibit the full range of qualitative behaviors (with selection acting primarily on heterozygotes in one and only on homozygotes in the other). Allele frequencies in the next generation follow from Wright-Fisher sampling with these viabilities, sometimes with migration, and the population size and migration rates vary according to the demographic scenario considered. We assume that fitness is multiplicative across sites and that selected sites are at Linkage Equilibrium and so each site's dynamics are independent from all other sites.

**Demographic scenarios.** We consider three demographic scenarios. The most detailed is the Out-of-Africa demographic model for African-Americans (AA) and European-Americans (EA) estimated by Tennessen et al.[6] (supplement Figure 1A).

The model includes the Out-of-Africa split of European ancestors, changes in population size before and after the split (specifically a severe bottleneck in Europeans following the split and recent rapid growth in both Europeans and Africans) and migration between the populations after the split. Finally, the model includes recent admixture between the populations, which we include in our simulations only when we compare our results to data from AAs.

We also study two simpler demographic scenarios. To understand the effects of recent explosive growth of human populations, we use a simple model of exponential growth from a population of constant size and similarly, to investigate the effects of the bottleneck in Europeans at the Out-of-Africa split, we consider a simple model of a bottleneck where population size instantaneously changes to a lower value in which it remains until it instantaneously reverts back to its original size.

**Simulations.** For each demographic scenario, we ran simulations of a single site for the semi-dominant and recessive cases and varied the selection coefficient such that selection ranges from effectively neutral to strong. Each run begins with one of the two alleles fixed, where the proportion of runs that start with each allele is given by the expectation at equilibrium. A burn-in period of $\geq 10N$ generations with constant population size $N$ follows in order to ensure an equilibrium distribution of segregating sites. The initial state is defined as ancestral and the other state as derived; the derived and deleterious allele frequencies are recorded at the end of the simulation. The code is written in C++ and is available upon request.

**Load.** Genetic load is defined as the relative reduction in average fitness caused by deleterious alleles [24]. Given our model, the average fitness function can be written as

$$\bar{W} \approx \exp(-\sum_{j=1}^{M} l(h_j, s_j))$$

where
$$l(h,s) \equiv 2hsE(pq) + sE(q^2) = s(2hE(q) + (1-2h)E(q^2)), \quad (1)$$

relates the quantities at a locus with load, $p$ and $q$ are the beneficial and deleterious allele frequencies at a locus ($p+q = 1$) and $h_j$ and $s_j$ are the dominance and selection coefficient at locus j. For a model with a single site and $s \ll 1$, $l(h,s)$ coincides with the definition of load. For more than one site, load is a simple function of the sum over $l(h,s)$'s. For brevity, we therefore refer to $l(h,s)$ as load.

**Change in load.** By changes in load, we refer to difference between load at the present and the load before recent demographic events. Specifically, in the exponential and bottleneck models the reference time is before the change in population size and in the OOA model the reference time is the split between the African and European populations.

**Data Analysis.** We used data from Fu et al. (2012) [7], who reported European-American (EA) and African-American (AA) allele frequency estimates and inferences of of the derived allele for each SNV in their dataset. . Variants for which allele frequencies are both 0 or both 1 in European- and African-Americans were excluded (presumably these are sites at which all sampled individuals differ from the reference genome sequence). Mean derived allele frequencies and their standard errors were calculated for both African Americans and European Americans at autosomal noncoding, synonymous, and nonsynonymous sites, as well as autosomal nonsynonymous variants belonging to the different functional categories.

The ANNOVAR suite of scripts [25] was used to obtain functional predictions for each SNP from each of four prediction methods: PolyPhen2 [21], SIFT [26], LRT [27] and MutationTaster [28]. Coding versus non-coding and synonymous versus nonsynonymous designations were also determined using ANNOVAR. Since mean allele frequencies of segregating sites are affected by total sample size, we subsampled the EA Autosomal and X and the AA Autosomal data to a sample size of 3852, which corresponds to the lowest sample size in the data, for AA X.

We observed a strong reference bias in PolyPhen2's functional classification: SNV's with the same derived allele frequency are classified as having greater functional severity when the reference genome carries the derived allele. We therefore treated the functional designations at these sites as unreliable. To correct for the exclusion of these sites, we binned SNVs by frequency and, for each bin, we estimated the fraction of sites in each functional category based on sites in which the reference genome carries the ancestral allele. Estimates of the derived allele frequencies in each function category were then calculated by weighting the contribution of each frequency according to these fractions.

**Models for variance.** We considered two extreme models for the relationship between effect sizes and selection coefficients: one in which they are independent and specifically, the effect size is constant and the other in which they are proportional to each other. The quantities in Figure 4 under these models were then calculated as follows. For simplicity, we assumed that mutations are semi-dominant for both fitness and the quantitative trait under consideration. At a site with selection coefficient $s$, the expected contribution to the variance from deleterious alleles below frequency $\omega$ is therefore

$$V_\omega(s) = \frac{1}{2}CE(a^2|s)\int_0^\omega f(x;s)x(1-x)dx, \tag{2}$$

where $E(a^2|s)$ is the expectation of the effect size squared for sites with selection coefficient $s$, $f(x;s)$ is the probability of the deleterious allele being at frequency $x$ (without conditioning of the site being segregating, i.e., including $x = 0$ and 1) and the $C$ is a proportion coefficient (cf. Supplement Section 4.1). The overall contribution to variance of a site is $V_1(s)$ and the fraction of that contribution coming from variants below frequency $\omega$ is $\Theta_\omega(s) \equiv \frac{V_\omega(s)}{V_1(s)}$; Note that while $V_1(s)$ depends on the relationship between selection coefficients and effect sizes, $\Theta_\omega(s)$ does not. When all sites are considered jointly, denoting the input of mutations with selection coefficient $s$ by $\mu(s)$, the expected proportion of variance from deleterious alleles below frequency $\omega$ is then

$$\Theta_\omega = \frac{\int_s \mu(s)V_1(s)\Theta_\omega(s)ds}{\int_s \mu(s)V_1(s)ds}. \tag{3}$$

As an illustration, we considered a simple example, in which a fraction $\alpha$ of the sites have a weak selection coefficient $s_w = 0.0002$ and a fraction $1 - \alpha$ have a strong selection coefficient of $s_s = 0.01$. In this case, the contribution of rare alleles (below $\omega = 0.1\%$) takes the form

$$\Theta_\omega(\alpha) = \frac{\alpha V_1(s_w)\Theta_\omega(s_w) + (1-\alpha)V_1(s_s)\Theta_\omega(s_s)}{\alpha V_1(s_w) + (1-\alpha)V_1(s_s)}. \tag{4}$$

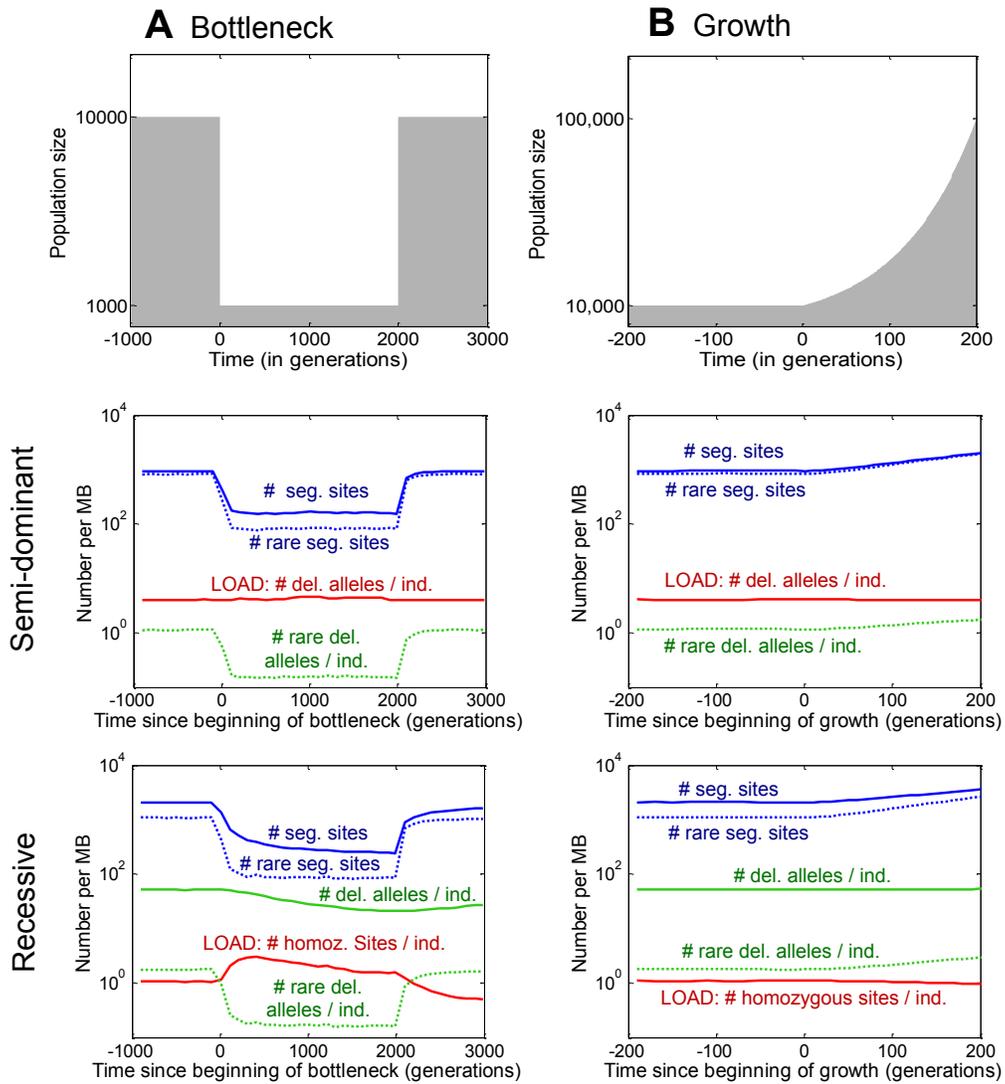

Figure 1: **Time course of load and other key aspects of variation through the course of a bottleneck (panels A, C, E) and exponential growth (panels B, D, F).** Each data line shows the expected number of variants, or alleles per MB, assuming semi-dominant mutations (panels **C**, **D**) or recessive mutations (panels **E**, **F**) with $s=1\%$ and mutation rate per site per generation$=10^{-8}$.

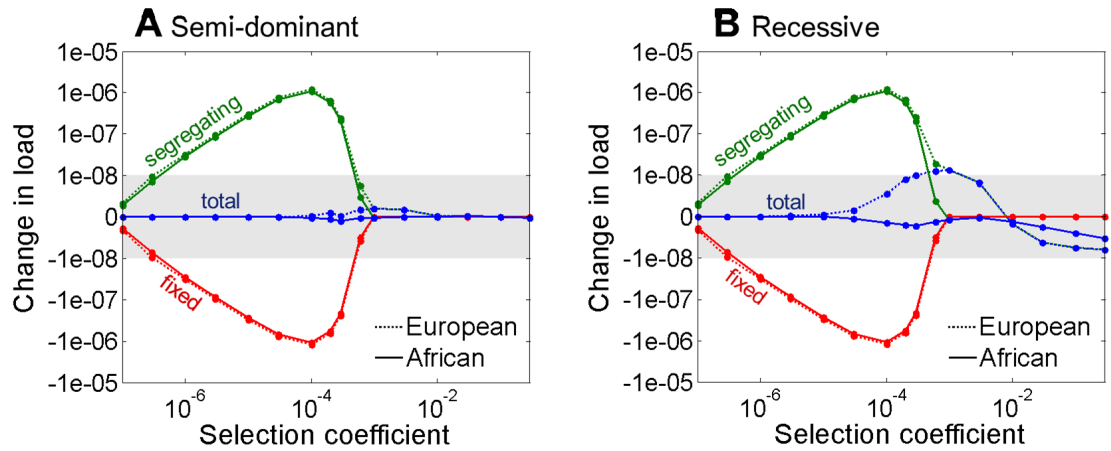

Figure 2: **Changes in load due to changes in population size during the histories of European and African Americans for (A) semi-dominant and (B) recessive sites.** The blue lines show the difference in total expected load per base pair of DNA sequence in the present day population compared to the ancestral (constant) population size, as a function of selection coefficient. The green and red lines show the difference in the amount of load due to segregating and fixed variants, respectively. As can be seen, there is more load due to segregating variation in modern populations, but this approximately cancels with reduced load to fixed sites as shown by the total load lines (blue). The y-axis scale is linear within the grey region and logarithmic outside.

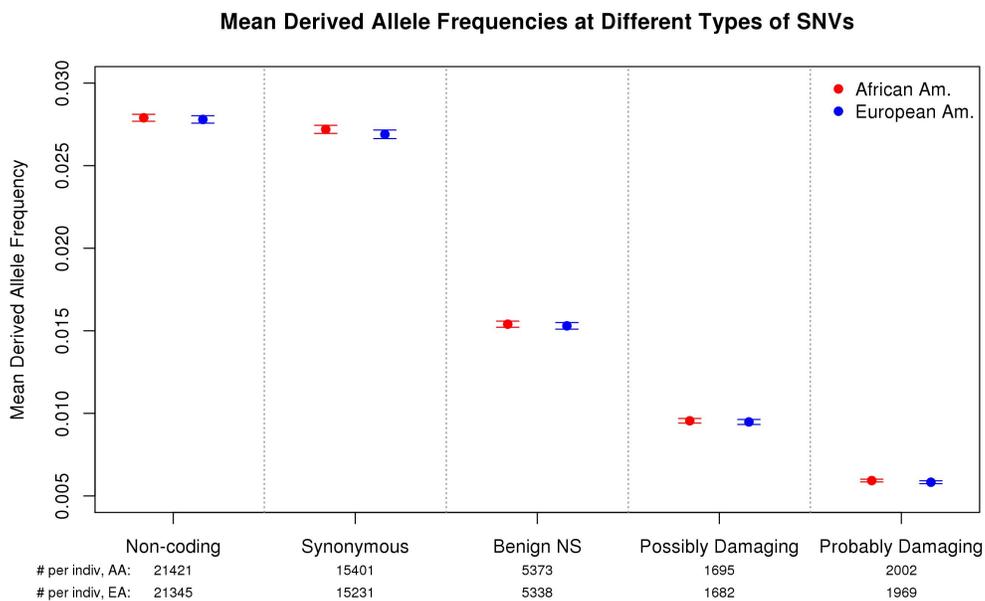

Figure 3: **Observed mean allele frequencies in African and European Americans at various classes of SNVs.** The plot shows mean frequencies in each population, plus and minus two standard errors, using exome sequence data from Fu et al.[7]. Here a site is considered an SNV if it is segregating in the combined AA-EA sample of 6515 individuals. The functional classifications of sites are from PolyPhen2[21] with modifications as described in the Supplement. The AA and EA mean frequencies are essentially identical within all five functional categories (p>0.05).

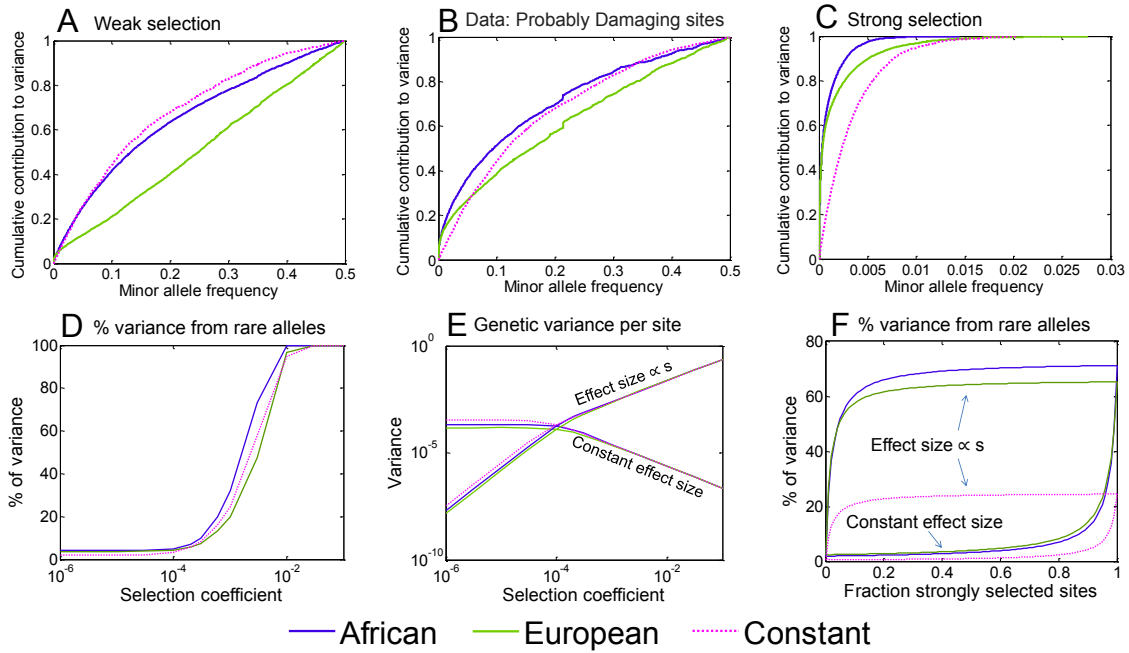

Figure 4: **Predicted contributions of different allele frequencies to variance in disease risk under the semi-dominant model.** The upper plots show the cumulative fractions of the total genetic variance that is due to alleles at frequency $< x$, for different models of selection and demography (note the differing x-axis ranges): (**A**) simulated data from the Tennessen *et al.*[6] demographic model for sites with weak selection ($s = .0002$); (**B**) assuming the observed frequency spectrum at "probably damaging" sites[7, 21]; (**C**) simulated data from the Tennessen *et al.*[6] demographic model for sites with strong selection ($s = .01$). Panel (**D**) shows the fraction of variance that is due to rare alleles (i.e., $< 0.1\%$ frequency) as a function of the selection coefficient; (**E**) shows the per-site contribution to variance as a function of the selection coefficient; (**F**) plots the fraction of the variance that is due to rare alleles if disease mutations are a mixture of strongly selected ($s = 0.01$) or weakly selected ($s = .0002$) sites. Panels E & F include results for two different models: either effect sizes are independent of $s$ (constant), or proportional to $s$. In all plots the purple dotted line shows simulation results for a constant-size model for comparison.

# Supporting Online Material for
# The deleterious mutation load is insensitive to recent population history

Yuval B. Simons, Michael C. Turchin, Jonathan K. Pritchard and Guy Sella

April 16, 2013

# Contents

# 1   Model and simulations

Our basic model considers selection at a single site. We use the standard bi-allelic diploid model with (in this order) two-way mutation, viability selection, drift and, in some cases, migration [1]. Specifically, we assume there are two alleles at a site: normal ($N$) and deleterious ($D$). An $N$ allele mutates to the $D$ allele with probability $u$ per gamete, per generation and the reverse mutation occurs with probability $v$. Unless noted otherwise, we assume that mutation is symmetric, i.e., $u = v$. The absolute fitnesses of the three genotypes $NN$, $ND$ and $DD$ are 1, $1 - hs$ and $1 - s$, respectively, where $s > 0$ and $h \geq 0$. We focus on semi-dominant ($h = \frac{1}{2}$) and fully recessive ($h = 0$) selection because these two cases exhibit the full range of qualitative behaviors (with selection acting primarily on heterozygotes in one and only on homozygotes in the other), but we also consider the robustness of our findings to other dominance coefficients (section 2.4). Allele frequencies in the next generation follow from Wright-Fisher sampling with these viabilities, sometimes with migration, and the population size and migration rates vary according to the demographic scenario considered.

For each demographic scenario, we ran simulations of a single site for the semi-dominant and recessive cases and varied the selection coefficient such that selection ranges from effectively neutral to strong. For a given set of parameters, the number of runs was determined by requiring a sampling error of less than 2% in estimates of the main summaries (e.g., the mean deleterious allele frequency and squared frequency). Error bars denoting estimates of one standard deviation around the mean are provided in all the graphs based on simulations, unless they are too small to be visible. Each run begins with one of the two alleles fixed, where the proportion of runs that start with each allele is given by the expectation at equilibrium. A burn-in period of $\geq 10N$ generations with constant population size $N$ follows in order to ensure an equilibrium distribution of segregating sites. The initial state is defined as ancestral and the other state as derived; the derived and deleterious allele frequen-cies are recorded at the end of the simulation. The code is written in C++ and is available upon request.

**Demographic scenarios.** We consider three demographic scenarios. The most detailed is the Out-of-Africa demographic model for African-Americans (AA) and European-Americans (EA) estimated by Tennessen et al. [2] (Figure 1A). The model includes the Out-of-Africa split of European ancestors, changes in population size before and after the split (specifically a severe bottleneck in Europeans following the split and recent rapid growth in both Europeans and Africans) and migration between the populations after the split (see Figure 1A for details). Finally, the model includes recent admixture between the populations, which we include in our simulations only when we compare our results to data from AAs.

While the Tennessen et al. model was parameterized in a diffusion framework, i.e., in continuous time, Wright-Fisher simulations require discrete numbers of generations and individuals. We therefore divide the times by 25 years per generation (the generation time that Tennessen et al. assume) and round the number of individuals associated with any of the parameters (e.g., growth) to the nearest integer. We implement migration by sampling alleles from the local population with probability $1 - m$ and from the other population with probability $m$ each generation.

We also study two simpler demographic scenarios. To understand the effects of recent explosive growth of human populations, we use a simple model of exponential growth with parameters matching those of the African population in the Tennessen et al. model (see Figure 1B for details). For the purpose of analysis, this scenario is sometimes extended by adding a period with constant population size after growth ends. Similarly, to investigate the effects of the bottleneck in Europeans at the Out-of-Africa split, we consider a simple model of a bottleneck with parameters matching those of the European bottleneck in the Tennessen et al. model (see Figure 1C for details). Here, we sometimes extend the period after the reduction in population size to study longer-term equilibration to reduced population sizes.

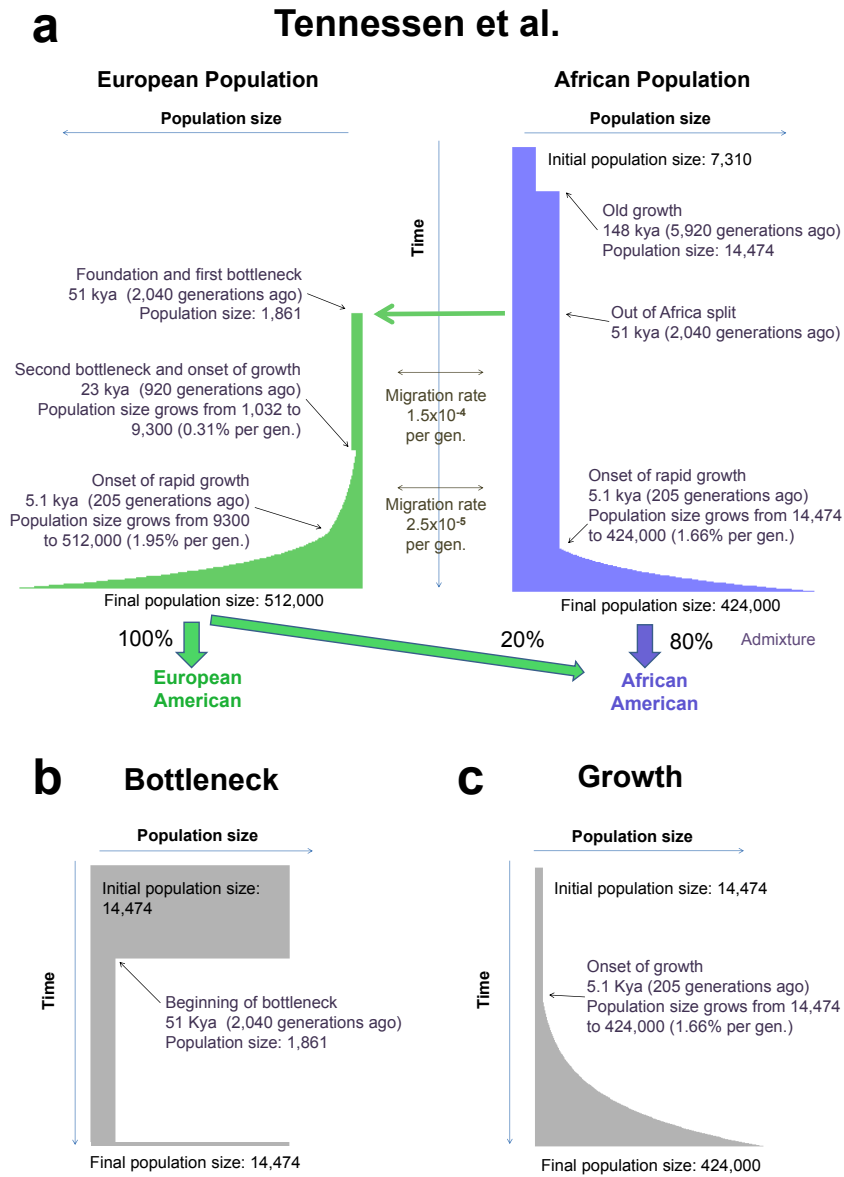

Figure 1: The three demographic models that we consider. A) The Out-of-Africa model estimated by Tennessen et al. [2]. B) A population bottleneck. C) Exponential growth. All population sizes are given as number of diploid individuals. In some cases, in order to study the equilibration process, we extend the bottleneck model to include a longer period with a reduced population size and the growth scenario to include a priod with a constant population size after growth.

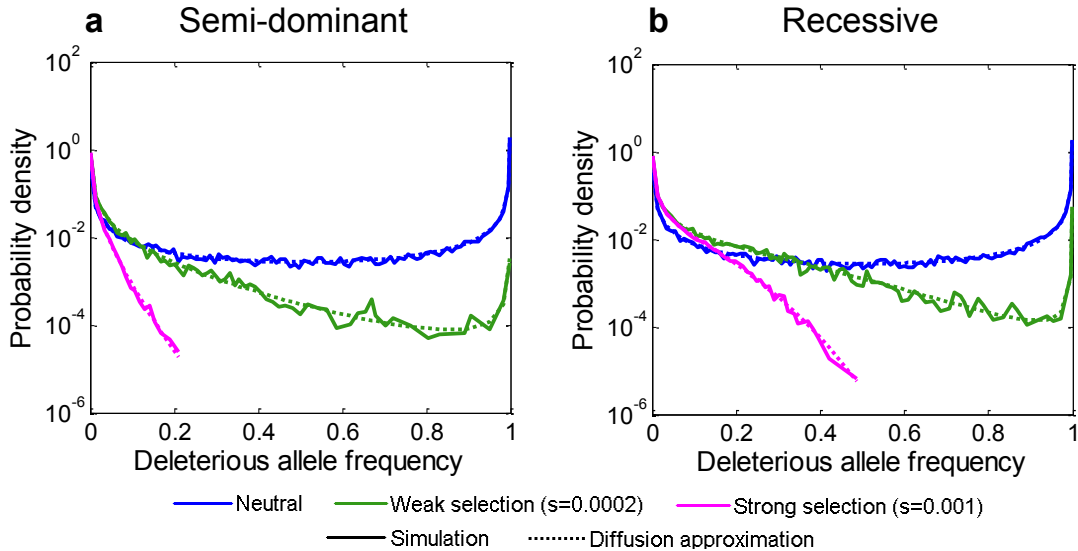

Figure 2: Comparison of theoretical and simulated frequency spectra for a constant population size in the (A) semi-dominant and (B) recessive models. Shown are the results based on the diffusion approximation (solid) and on simulations (dashed) for several selection coefficients. The population size was taken as $N = 14,474$ and the mutation rate as $u = 2.36 \cdot 10^{-8}$ per generation per site. The number of runs for each set of parameters was $10^6$.

**Checking the simulation.** We used two approaches to check the validity of the simulations. For a constant population size, we compared the frequency spectra from simulations with those expected under the diffusion approximation (cf. [3]) for the neutral case as well as for several semi-dominant and recessive selection coefficients (Figure 2). We note that obtaining similar frequency spectra implies that simpler summaries, such as the number of segregating sites under neutrality or the average deleterious allele frequency at mutation-selection balance, will also be similar.

For the more elaborate Out-of-Africa demographic model, we compared the minor allele frequency spectrum from neutral simulations with the spectrum observed at non-coding sites in Fu et al. [4] We cosider non-coding sites for this purpose as these are assumed to be under the least selection (Figure 3). In their Figure 2A, Tennessen et al. find a close agreement between the observed spectra and a diffusion approximation under their demographic model. We find close agreement of our neutral simulations to data from both AAs and EAs and the slight differences that we do find are similar to those in their Figure 2A [2].

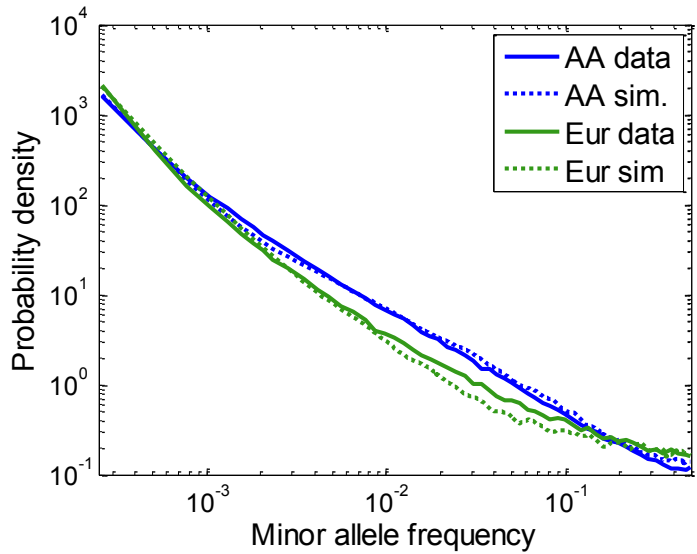

Figure 3: Comparison of the minor allele frequency spectrum in data from Fu et. al. and in simulations based on the Tennessen et al. model. The spectra are for a sample size of 3852 chromosomes in AA and EA populations, for both the data and simulations.

**Sensitivity to mutation rate.** Unless noted otherwise, we follow Tennessen et al. [2] in using a mutation rate of $u = 2.36 \cdot 10^{-8}$ per bp per generation. Given that recent estimates suggest a lower mutation rate (e.g. Kong et al. [5], Sun et al. [6]), we examine here the sensitivity of our simulation results to this assumption. We find the derived allele frequency spectrum to be extremely robust, remaining essentially unchanged when we double or halve the mutation rate (Figure 4A). As expected, the number of segregating sites and the number of sites fixed for the derived allele increase (linearly) with the mutation rate (Figure 4B). The increase in the number of sites fixed for the derived allele follows from the increased rate of fixation in the burn in period (akin to fixations that occur between the ancestor of humans and chimpanzees and the Out-of-Africa split). Thus, assuming a different mutation rate will affect some of our quantitative results. Notably, if the mutation rate in humans is indeed lower than the one we use, as recent estimates suggest, the proportion of segregating sites would be lower, resulting in an even smaller effect of recent demographic history on load than our analysis suggests (see section 2). Our qualitative finding of a negligible effect on load is unchanged. Moreover, our results concerning the effects of recent demography on genetic architecture derive from the frequency spectrum and therefore are unaffected.

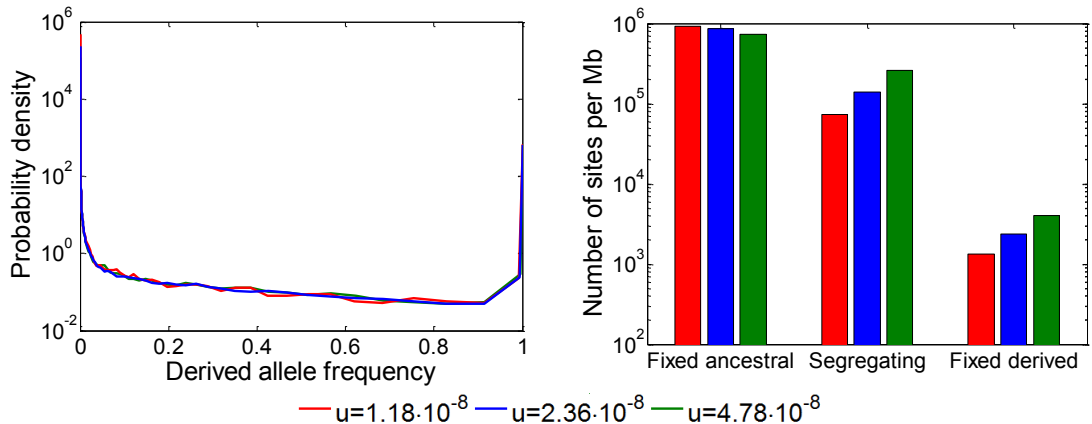

Figure 4: Sensitivity of (A) the frequency spectrum and (B) the number of segregating and fixed sites to the mutation rate. The results are shown for simulations of the African population but are qualitatively similar for the European population.

## 2  The effects of demography on load

We assume that fitness is multiplicative across sites and that selected sites are at Linkage Equilibrium (LE). The absolute fitness of individual $i$ can then be written as

$$W_i = \prod_{j=1}^{M} w_{i,j},$$

where the product is taken over the $M$ sites contributing to fitness and $w_{i,j}$ is the contribution of site $j$, which depends on the genotype of the individual and on the selection and dominance coefficients at that site. Given LE, the contributions of sites to the expected fitness in the population are independent and therefore

$$E(W_i) = \prod_{j=1}^{M} E(w_{i,j}) \approx \exp(-\sum_{j=1}^{M}(2h_j s_j p_j q_j + s_j q_j^2)),$$

where $p_j$ and $q_j$ are the frequencies of the normal and deleterious alleles at locus $j$. We note that the approximation applies for strong selection because the frequency $q_j$ is small, as well as for weak selection because then the selection coefficient is small. Finally, taking an expectation over evolutionary realizations (which is equivalent to an expectation over many sites with the same parameters in a single realization) yields

$$E(W) \approx \exp(-\sum_{j=1}^{M}(2h_j s_j E(p_j q_j) + s_j E(q_j^2))). \qquad (1)$$

The latter expression relates the population dynamics at a site with the overall reduction in fitness.

Genetic load is defined as the relative reduction in average fitness caused by deleterious alleles, calculated as

$$L = \frac{W_{max} - \bar{W}}{W_{max}},$$

where $W_{max}$ is the fitness of an individual without deleterious alleles and $\bar{W}$ is the average fitness [1]. Denoting the terms associated with a single site in Equation 1

by
$$l(h, s) \equiv 2hsE(pq) + sE(q^2) = s(2hE(q) + (1 - 2h)E(q^2)), \qquad (2)$$

the fitness function can be rewritten as

$$E(W) \approx \exp(-\sum_{j=1}^{M} l(h_j, s_j)).$$

This form emphasizes that the reduction in fitness caused by a single site generally depends on the first two moments of the deleterious allele frequency. Specifically, in the semi-dominant model, it depends only on the first moment

$$l(\frac{1}{2}, s) = sE(q),$$

and in the recessive model it depends only on the second

$$l(0, s) = sE(q^2).$$

Moreover, this form shows that $l(h, s)$ provides a natural additive measure for the expected reduction in fitness caused by a site.

Throughout the manuscript we therefore use $l(h, s)$ as our measure for the contribution of a site to load. For a model with a single site, it coincides with the definition of load, as $E(L) = l(h, s)$. For more than one site,

$$E(L) \approx 1 - Exp(-\sum_{j=1}^{M} l(h_j, s_j)).$$

Given that in our model, the load from all sites is a simple function of the sum of $l(h, s)$ across sites, for brevity, we refer to $l(h, s)$ as load.

With a constant population size, the load exhibits three standard dynamic regimes depending on the scaled selection coefficient (Figure 5): (i) An effectively neutral regime, in which $\alpha = 2Ns \ll 1$ and the effects of selection are negligible compared to drift; (ii) a weak selection (or nearly neutral) regime, in which $\alpha = 2Ns \approx 1$ and the effects of selection and drift are comparable; (iii) a strong selection regime, in which $\alpha = 2Ns \gg 1$ and selection dominates over drift.

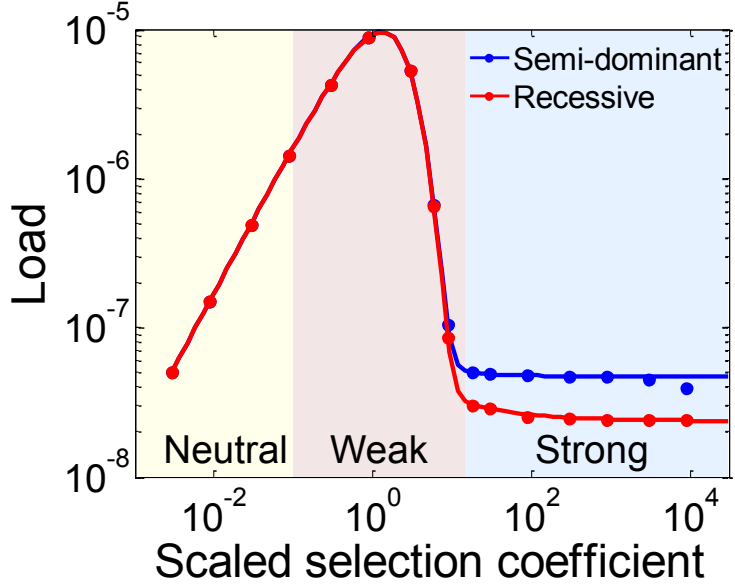

Figure 5: Load as a function of selection coefficient in a population of constant size. Results are shown for the semi-dominant (blue) and recessive models (red), where the diffusion approximation is shown as a solid line and simulation results as circles. The population size is $N = 14,474$.

In what follows our analysis is divided according to these three regimes. When the population size changes, the boundaries between regimes are affected. Moreover, the rate at which the equilibrium for a new population size is attained depends on the summary of the data considered. We consider summaries for segregating sites, e.g., the proportion of segregating sites and the allele frequency at these sites, and summaries for fixed sites, e.g., the proportion of sites fixed for the deleterious allele (which we call fixed state). Specifically, we are interested in the effects of demography on the contribution of segregating and fixed sites to load, which we refer to as fixed and segregating load, and in their sum, which we refer to as total load. We consider the behavior of these statistics for the two simple demographic models, which together allow us to understand all qualitative behaviors exhibited under the more detailed Tennessen et al. model (6). For these demographic models, we primarily consider two modes of inheritance (semi-dominant and recessive).

To simplify our theoretical analysis, we make several reasonable assumptions about the parameters of the model. For brevity, we focus on the case with symmetric mutation ($u = v$) and, because we are considering human populations, we assume that the population mutation rate per site is small, i.e., that $\beta = 2Nu \ll 1$. We also assume that the selection coefficient is small, i.e., $s \ll 1$. A summary of our analyses are presented in Figure 6 and Table 1. A detailed description of the behavior in each regime follows.

|  |  |  | Effectively neutral | Weak | | Strong |
|---|---|---|---|---|---|---|
|  |  |  |  | closer to neutral | closer to strong |  |
| Bottleneck | Semi-dominant | fixed | increase | increase | increase | — |
|  |  | segregating | decrease | decrease | increase | unchanged |
|  |  | total | unchanged | increase | increase | unchanged |
|  | Recessive | fixed | increase | increase | increase | — |
|  |  | segregating | decrease | decrease | increase | transient increase |
|  |  | total | unchanged | increase | increase | transient increase |
| Growth | Semi-dominant | fixed | decrease | decrease | | — |
|  |  | segregating | increase | increase | | unchanged |
|  |  | total | unchanged | unchanged | | unchanged |
|  | Recessive | fixed | decrease | decrease | | — |
|  |  | segregating | increase | increase | | transient decrease |
|  |  | total | unchanged | unchanged | | transient decrease |

Table 1: Changes to load under the bottleneck and growth models. The effects on fixed, segregating and total load are depicted by selection regime. The symbol — denotes the cases in which there is no contribution to load both before and after the change in population size.

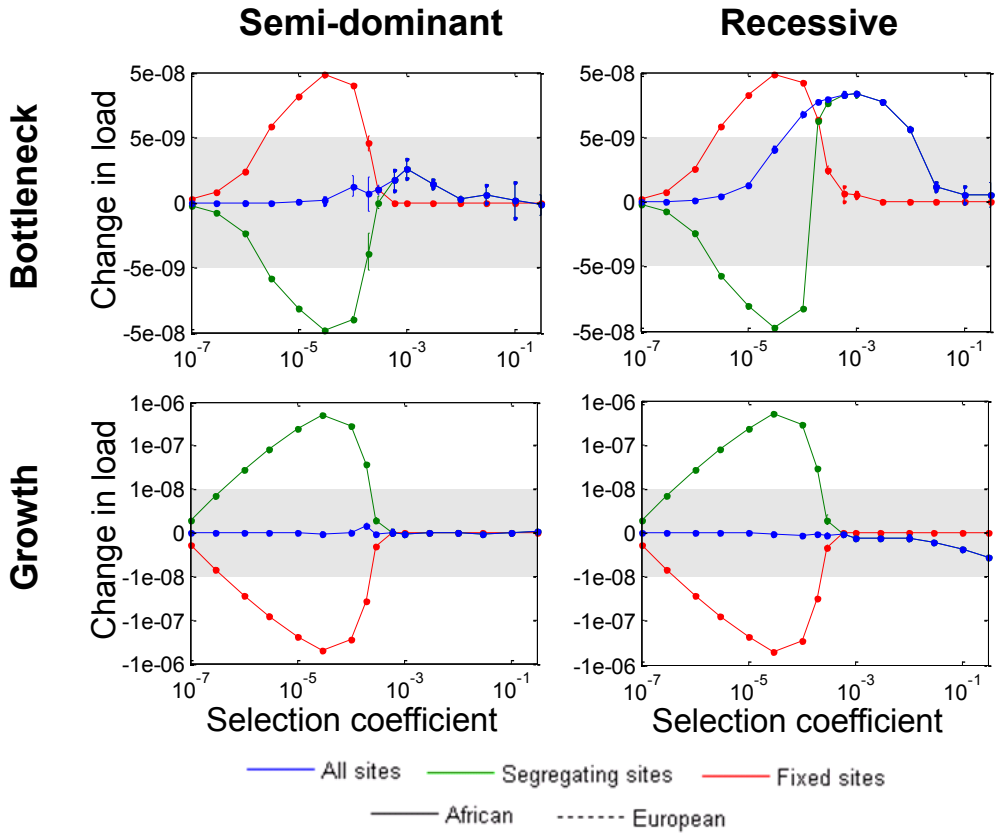

Figure 6: The changes to the segregating, fixed and total load under the bottleneck and growth models. Analogous graphs for the Tennessen et al. model are presented in Figure 3 of the main text. Changes are measured by comparison to a population in which the population size has remained constant at the size that it was at the beginning of the demographic model. In the shaded areas, load is shown on linear scale; otherwise it is shown on logarithmic scale.

## 2.1 The effectively neutral regime

When selection is negligible compared to drift, the behavior of deleterious alleles is well approximated by that of neutral alleles. As the properties of neutral alleles (e.g., the proportion of segregating sites and frequency spectrum) in models with constant and varying population sizes have been studied exhaustively (e.g., [8, 9, 10]), here we focus only on the implications concerning load.

First, we consider how load depends on the selection coefficient at equilibrium for a constant population size. If deleterious alleles behave like neutral ones, the first two moments of the deleterious allele frequency distribution do not depend on the selection coefficient and therefore the load is proportional to the selection coefficient (see Eq. 2). This explains the linear relationship between selection coefficient and load shown in Figure 5.

At equilibrium, load depends negligibly on the population size. Using the diffusion approximation for the stationary deleterious allele frequency distribution [3], the expansion of the load to first order in $\alpha$ and $\beta$ yields

$$l(h, s) = \frac{s}{2}(1 - \frac{1}{2}\alpha - 2(1-2h)\beta).$$

Thus, as long as $\beta \ll 1$ and $\alpha \ll 1$, the load is well approximated by $s/2$ regardless of the population size and dominance coefficient (hence the similarity in load for the semi-dominant and recessive cases in Figure 5). Intuitively, this follows from the fact that the great majority of sites are fixed, and because selection is negligible, half of them are fixed for the deleterious allele ($\frac{u}{u+v}$ for asymmetric mutation).

The same reasoning implies that changes in population size will have a negligible effect on the total load in this regime (Figure 7). While changes in population size affect the proportion of segregating sites and thus their contribution to load, so long as the population mutation rate remains negligibly small ($\beta \ll 1$), the segregating load will remain negligible compared to the fixed load. In the bottleneck model, the proportion of segregating sites decreases to a new equilibrium after the reduction in population size (Figure 7A). This explains the decrease in segregating load, which is balanced by an increase in fixed load (Figure 6). By the same token, in the growth model, the segregating load increases but is balanced by a decrease in fixed load, resulting in a negligible change to the total load (Figure 6 and Figure 7B). In this case, however, segregating sites are still far from their new equilibrium at present (see the next section).

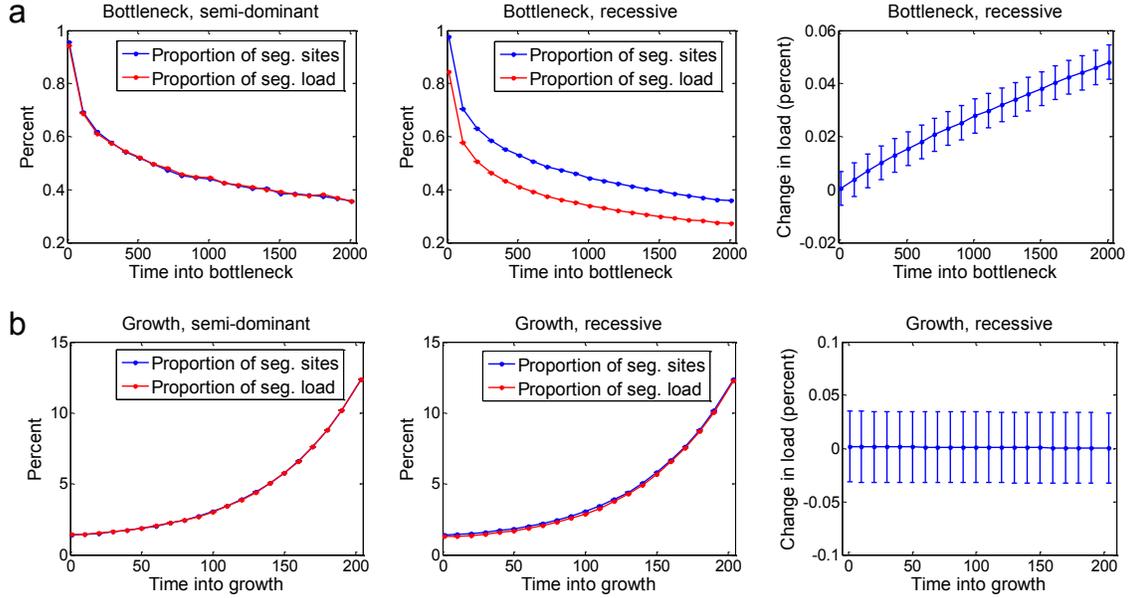

Figure 7: Segregating and total load in the bottleneck and growth models in the effectively neutral regime. The proportion of segregating sites, their proportional contribution to load, and the proportional change in total load are shown as a function of time (A) after the bottleneck and (B) since the onset of growth. The selection coefficient is $s = 10^{-7}$. In the semi-dominant case, the expected total load is always $s/2$ regardless of changes in population size; in the recessive case, changes to the proportion of segregating sites affect the total load, but this effect is negligibly small.

## 2.2 The weak selection regime

In the weakly selected regime, selection and drift have comparable effects on the dynamics of deleterious alleles. As a result, at equilibrium, even moderate differences in population size can affect the balance between selection and drift. Changes in population size also shift the balance, and are followed by transient changes at fixed and segregating sites until a new equilibrium is attained. To understand these effects, we consider the behavior at equilibrium and the rate at which it is approached. For this purpose, it is helpful to use the low mutation rate (LMR) approximation in which mutant alleles at a segregating site have a single origin; in other words, we ignore mutations that arise during the sojourn of a mutant allele from the time it arises on a background fixed for the other allele to the time it reaches fixation or loss in the population.

**The effect of population size on the proportion of sites fixed for the normal and deleterious alleles.** At equilibrium, the rate at which deleterious alleles arise and fix is equal to the rate at which normal alleles arise and fix. This balance can be written as

$$2Nup\pi(-2Ns, h, \frac{1}{2N}) = 2Nvq\pi(2Ns, 1-h, \frac{1}{2N}),$$

where $\pi$ denotes the fixation probability, which depends on the scaled selection and dominance coefficients and on the initial frequency [11] (because $s \ll 1$, we ignore second order terms in $s$). For $s \ll 1$ and any dominance coefficient, this yields

$$\frac{q}{p} = \frac{u}{v} \frac{\pi(-2Ns, h, \frac{1}{2N})}{\pi(2Ns, 1-h, \frac{1}{2N})} \approx \frac{u}{v} e^{-2Ns}.$$

Namely, at equilibrium, the proportion of fixed deleterious sites declines exponentially with the scaled selection coefficient $\alpha = 2Ns$ (Figure 8A). Thus, for a given selection coefficient $s$, the population size has a dramatic effect on the proportion of sites fixed for the deleterious allele, declining from the neutral, mutation-driven, proportions for $s \ll \frac{1}{2N}$ to approximately 0 for $s \gg \frac{1}{2N}$.

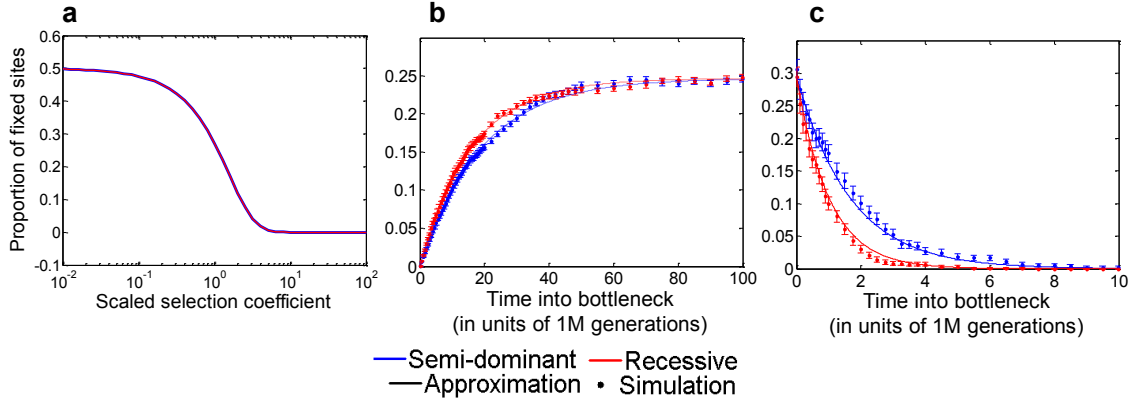

Figure 8: Proportion of sites fixed for deleterious alleles in the weak selection regime. In all graphs, the selection coefficient is $s = 10^{-4}$. (A) The equilibrium proportion as a function of the scaled selection coefficient ($\alpha = 2Ns$), where the population size was varied. (B) The proportion as a function of time after the change in population size in the bottleneck model. (C) The proportion as a function of time after the change in population size in the growth model.

Importantly, however, when the population size changes, the new equilibrium proportion may be attained very slowly. The fractions, $p(t)$ and $q(t)$, of sites fixed for the normal and deleterious alleles $t$ generations after a change in population size (assuming $p(t) + q(t) = 1$) are well approximated by the model

$$\frac{\mathrm{d}}{\mathrm{d}t} \begin{pmatrix} p \\ q \end{pmatrix} = \begin{pmatrix} -2N_a u \pi(-2N_a s, h, \frac{1}{2N_a}) & 2N_a v \pi(2N_a s, 1-h, \frac{1}{2N_a}) \\ 2N_a u \pi(-2N_a s, h, \frac{1}{2N_a}) & -2N_a v \pi(2N_a s, 1-h, \frac{1}{2N_a}) \end{pmatrix} \begin{pmatrix} p \\ q \end{pmatrix},$$

where $N_a$ is the population size after the change, and fixation times (on the order of $4N_a$ generations) are neglected. An additional contribution from sites that were segregating before the change is considered below. In this approximation, the change in the fraction of sites fixed for the deleterious alleles is

$$q(t) = q_a^{eq}\left(1 - e^{-\frac{t}{\tau}}\right) + q_b^{eq} e^{-\frac{t}{\tau}},$$

where $q_b^{eq}$ and $q_a^{eq}$ are the equilibrium fractions corresponding to the population sizes before and after the change, and

$$\tau = \left[2N_a\left(u\pi(-2N_a s, h, \frac{1}{2N_a}) + v\pi(2N_a s, 1-h, \frac{1}{2N_a})\right)\right]^{-1}$$

is the timescale of the exponential approach to the new equilibrium. For the semi-dominant case and $s \ll 1$, this time scale is well approximated by

$$\tau \approx \left[u\frac{\alpha}{e^\alpha - 1} + v\frac{\alpha}{1 - e^{-\alpha}}\right]^{-1},$$

demonstrating that it is mutation-limited. This is also true for other dominance coefficients. In other words, following an instantaneous change in population size, the proportion of sites fixed for the deleterious allele will change extremely slowly, at a rate that is inversely proportional to the mutation rate (Figure 8B and C).

Because the equilibrium is reached slowly, recent demographic changes in humans should have had little effect on the proportion of sites fixed for the deleterious alleles and hence on the fixed load. The bottleneck at the Out-of-Africa split is estimated to have reduced the population size from $\sim 14,000$ to $1,800$ approximately 2000 generations ago [2]. Once a new equilibrium is reached, there will be a substantial increase in the proportion of fixed deleterious alleles; for example, for a semi-dominant deleterious allele with selection coefficient of $s = 10^{-4}$, it would increase it from 0.05 to 0.4. Yet the change over 2000 generations is minimal, increasing this proportion only by $3 \cdot 10^{-5}$. The estimated 200 generations since the onset of rapid growth in humans is similarly much too short a time period for any measurable effect on the fixed load (which in this case would decrease over large time periods).

**The effects of population size on segregating sites.** First we consider how the equilibrium properties of segregating sites depend on population size in models with constant population size (Figure 9). The deleterious allele frequency at segregating sites decreases with increasing population size, because the efficacy of selection is greater in larger populations (Figure 9A). In turn, the proportion of segregating sites increases with population size due to the (linear) increase in the number of mutations that enter the population every generation (Figure 9B). This is true not only for the population as a whole but also for subsamples from it of any size (Figure 9C). Finally, the deleterious allele frequency and proportion of segregating sites decrease with increasing dominance coefficient, as stronger selection in heterozygotes results in stronger selection on deleterious mutations (regardless of their frequency) and thus in a shorter sojourn through the population. Thus, in larger populations or if the dominance coefficient is greater, we expect a greater proportion of segregating sites with deleterious alleles at lower frequency.

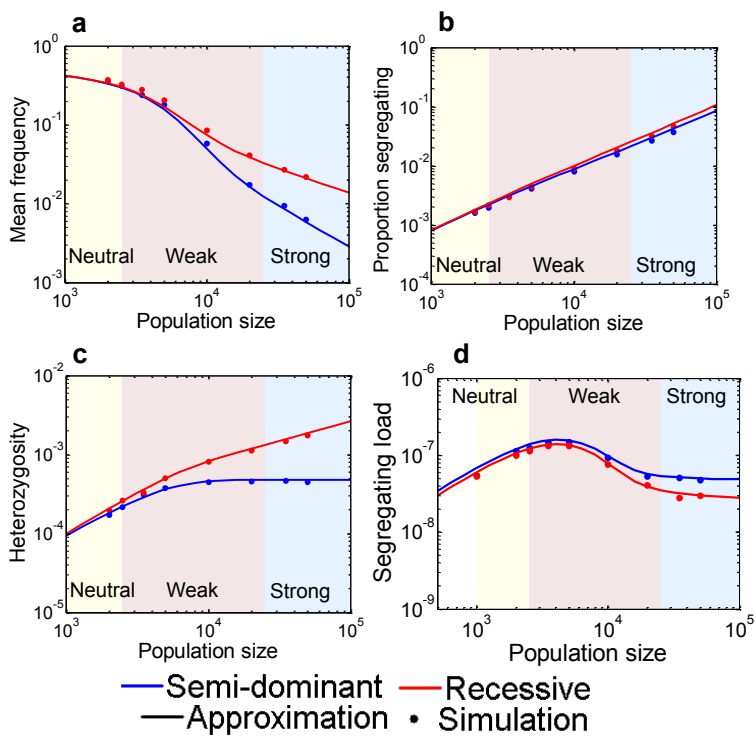

Figure 9: Equilibrium properties of segregating sites as a function of population size in constant population size models. In all graphs, $s = 2 \cdot 10^{-4}$. (A) The average frequency of segregating deleterious alleles. (B) The proportion of segregating sites. (C) Heterozygosity. (D) Segregating load.

The total load decreases monotonically when the population size increases (as can be shown using the stationary distribution based on the diffusion approximation [3], for example). This is not true of the segregating load, because the increase in the mutational input can have a greater effect than the increase in the efficacy of selection (Figure 9D). Indeed, for selection coefficients closer to neutrality, the increase in mutational input (and the proportion of segregating sites) dominates, causing the segregating load to increase with population size (akin to the behavior in the effectively neutral regime). In contrast, for selection coefficients closer to the strong selection regime, the increase in the efficacy of selection dominates, leading to a reduction in segregating load (akin to the stronger selection regime; see section 2.3).

Next we consider the effects of a change in population size. We begin by noting that, for a given population size, the expected sojourn time of deleterious and beneficial mutations that reach fixation is shorter than that for a neutral mutation and is thus on the order of $4N$ generations or less [3]. This implies that on the order of $4N_a$ generations after a change in population size, most of the *old mutations* (i.e.., those that segregated before the population size changed) have been absorbed (either due to loss or fixation), and replenished by *new mutations* (that arose and spread through the population at its new size). When this turnover process is complete, new segregating sites approach their equilibrium proportions (given a background of fixed sites).

In the bottleneck model, the reduction in the efficacy of selection causes an increase in total load, where the behavior of the components of load can be understood as follows (Figure 10). Focusing first on the contribution of old mutations to the fixed load: When old mutations are absorbed, the reduction in the efficacy of selection leads more deleterious alleles to fix than would have had the population size remained constant (at the larger size), eventually resulting in an increase in fixed load. The increase can be approximated by

$$\Delta(s, h, u, N_b, N_a) = \int_0^1 \left(\pi(-2N_a s, h, x) - \pi(-2N_b s, h, x)\right) f(x; h, 2N_b s, 2N_b u) dx,$$

where $f(x; h, 2N_b s, 2N_b u)$ is the stationary distribution before the change in popula-tion size [3]. The increase is maximized for selection coefficients at which the change in population size leads selection to transit from strong to weak, and is negligible outside this range (Figure 10A; explaining why it is more pronounced in Figure 10C and D than in E and F, correspondingly). The increase in deleterious fixations and load is then followed by a long-term, slower increase in the fixed load due to new mutations (Figure 10C-F). In the parameter regime where the fixation of old mutations makes a substantial contribution to load, there is also a transient increase in segregating load before the mutations fix (in Figure 10C for example). These effects are more pronounced in the recessive case, because of the greater frequency and proportion of segregating sites. Now focusing on the segregating load (Figure 10B): when segregating sites attain equilibrium, the reduction in population size causes a decrease in segregating load for lower selection coefficients (Figure 10C and D) and an increase for higher selection coefficients (Figure 10E and F). Thus, for higher selection coefficients in the weak selection range, both old and new mutations contribute to the transient increase in segregating load observed in Figure 6. For the lower selection coefficients in this range, the segregating load decreases both in the short and long term but the fixation of old mutations still results in an overall increase to the total load (Figure 6). Importantly, however, on the timescale estimated for the bottleneck at the Out-of-Africa split (vertical line in Figure 10), these effects amount to a tiny increase in total load (Figure 6).

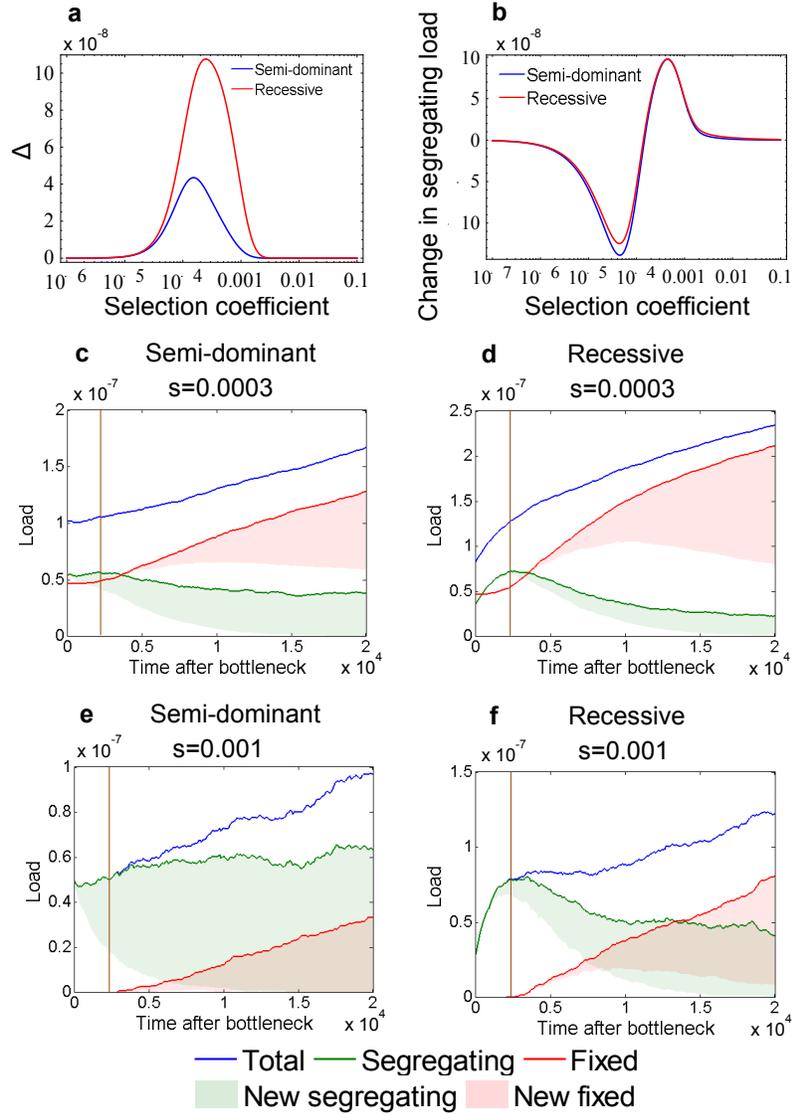

Figure 10: The changes in load shortly after a bottleneck. The figure shows (A) the expected change in fixed load due to mutations that segregated before the bottleneck and (B) the expected change in segregating load due to the bottleneck as a function of the selection coefficient. Shown are segregating, fixed and total load from new and all mutations as a function of time since the population size decrease. The (C and E) semi-dominant and (D and F) recessive cases are shown with a selection coefficient in the weak selection regime closer to neutral ($s = 0.0003$) and closer to strong ($s = 0.001$).

What about in the case of growth? Human population growth is thought to have started a couple hundred of generations ago, ending with an effective population size in the hundreds of thousands and starting from a size that was thirty-fold smaller [2]. Given the estimated growth parameters, there was insufficient time for the deleterious alleles that segregated before the onset of growth to change their frequencies substantially. Indeed even with the increase in the efficacy of selection as the population size increases, in this regime, selection is too weak to have caused a substantial change in allele frequency over hundreds of generations (although it could have caused the absorption of very rare or very high frequency alleles). After growth, the resulting frequency spectrum of deleterious alleles thus reflects a superposition of the spectrum of segregating sites before growth and of the spectrum at the large number of sites in which mutations were introduced after the onset of growth (Figure 11). The many new mutations remain at low frequencies. Because of an increase in the proportion of segregating sites, the segregating load increases at the expense of fixed load, but with negligible effects on the total load, given both the low frequency of new mutations as well as the opposing contributions of normal and deleterious mutations (Figure 6).

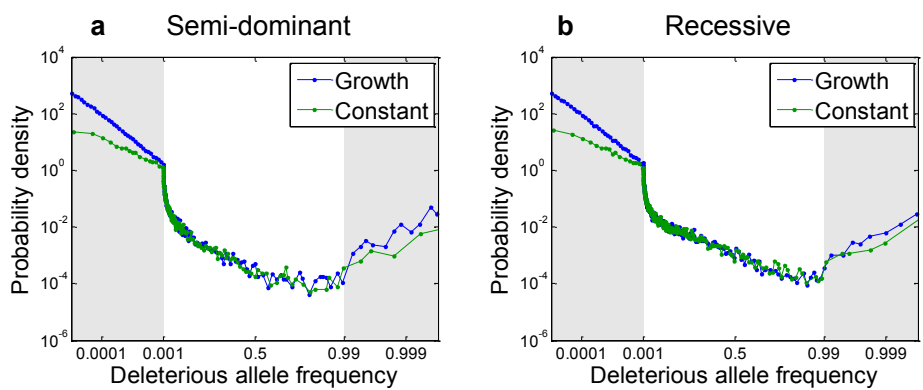

Figure 11: The frequency spectrum of weakly deleterious segregating sites in models with and without growth. In the shaded areas, frequency is shown on logarithmic scale; otherwise it is shown on linear scale.

## 2.3 The strong selection regime

In this regime, purifying selection is sufficiently strong to prevent deleterious alleles from reaching high frequencies, let alone fixation. It follows that there is only segregating load. If we assume that the deleterious allele frequency is small and that the dominance coefficient is sufficiently large, then the load is well approximated by

$$l(h, s) \approx 2hsE(q).$$

Stated another way, when selection against heterozygotes is sufficiently strong, then deleterious homozygotes would be too rare to affect load. Under these assumptions, the diffusion approximation at equilibrium with a constant population size [3] yields

$$E(q) \approx \frac{u}{hs},$$

implying that the load is well approximated by

$$l(h, s) \approx 2u.$$

We refer to the cases where these conditions are met as quasi-dominant.

In the recessive case, the load depends on the second moment of deleterious allele frequency. Assuming once again that the deleterious allele frequency is small, the diffusion approximation at equilibrium with a constant population size [3] yields

$$E(q^2) \approx \frac{u}{s},$$

implying that the load is well approximated by

$$l(0, s) \approx u.$$

The expressions for load in both cases are identical to the classic ones for mutation-selection balance, which are derived assuming an infinite population size [11]. They imply that at equilibrium, the load depends neither on the selection coefficient (explaining the plateaus in Figure 5) nor on the population size.

When the dominance coefficient is sufficiently small, however, the load does depend on population size (Figure 12). This will be the case when selection against heterozygotes is weak, i.e. when $2Nhs \gg 1$ does not hold, as then both moments of deleterious allele frequency make comparable contributions to load. Holding the selection coefficient and population size constant, in this range of dominance coefficients, the load varies continuously with $h$ between $u$ and $2u$ (Figure 12A). In turn, holding $h \ll 1$ and $2Ns \gg 1$ constant, increasing $s$ also leads the load to vary from $u$ to $2u$ (Figure 12B).

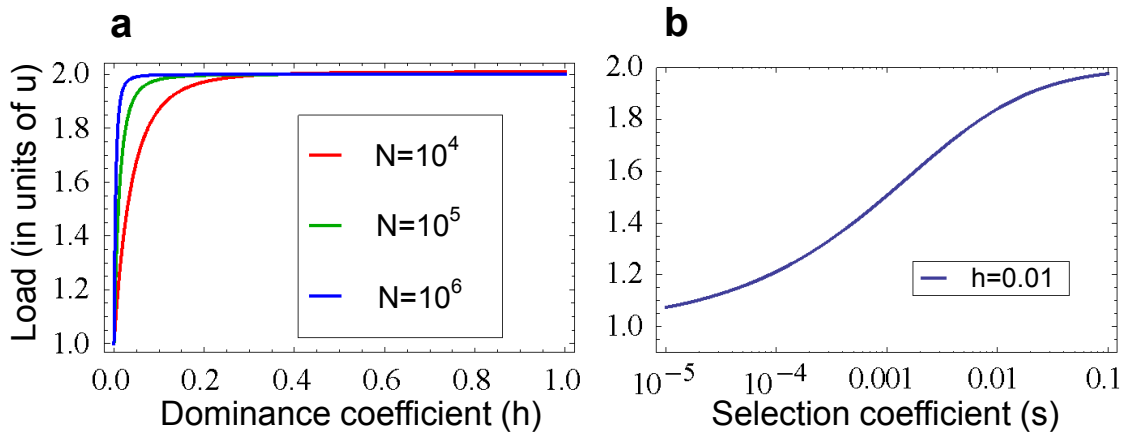

Figure 12: The dependence of the load on the dominance coefficient at equilibrium. The graphs were generated using the diffusion approximation for the stationary distribution assuming that the deleterious allele frequency is small [3]. A) Load as a function of the dominance coefficient $h$, with $s = 0.01$ and population size $N = 10^4, 10^5$ and $10^6$. B) Load as a function of the selection coefficient $s$, with $h = 0.01$ and $N = 10^6$.

Next, we consider the effect of changes in population size, for the quasi-dominant and then the recessive case. We show that in the quasi-dominant case, the load remains constant and is well approximated by the classic derivations for mutation-selection balance. In the recessive case, the load exhibits transient changes before it returns to its equilibrium level.

**The quasi-dominant case**

In the quasi-dominant case, we can assume deleterious alleles are sufficiently rare that selection against deleterious homozygotes can be ignored and selection has negligible effects on average fitness. Under these conditions, we can approximate the trajectory of a deleterious allele using a branching process (cf. [12]), in which the number of copies that a given deleterious allele gives rise to in the next generation follows a distribution that is independent on the frequency of deleterious alleles in the population.

Consider a single deleterious allele that was introduced by mutation at time $t = 0$ and denote by $Z(t)$ the number of deleterious alleles that it gives rise to at generation $t$. The number of mutant alleles in the next generation can then be expressed as

$$Z(t+1) = \sum_{i=1}^{Z(t)} X_i(t),$$

where $X_i(t)$ denotes the number of offspring of the $i$th allele at time $t$ and $i = 1, \ldots, Z(t)$. We denote the expected number of offspring of a single allele by $\lambda$, i.e., $E(X_i(t)) = \lambda$; if we ignore mutations back to the beneficial allele then $\lambda = 1 - hs$ and if we include them then $\lambda = 1 - hs - v$. The expected number of alleles in the next generation is then

$$E(Z(t+1)) = E(\sum_{i=1}^{Z(t)} X_i(t)) = \sum_{j=1}^{\infty} Pr(Z(t) = j) j E(X_i(t)) = E(Z(t))\lambda, \quad (3)$$

or

$$E(Z(t)) = \lambda^\tau. \quad (4)$$

Now consider the expected number of deleterious alleles at mutation-selection balance. For this purpose, we measure time backwards from the present. We denote by $Y_\tau(\tau)$ the number of mutations introduced $\tau$ generations ago and by $Y_\tau(t)$ the number of alleles that they give rise to at time $t$. The number of deleterious alleles at the present can then be expressed as the sum of contributions from all the mutations in the past, i.e. $\sum_{\tau=1}^{\infty} Y_\tau(0)$, where, from Equation 4,

$$E(Y_\tau(0)) = Y_\tau(\tau)\lambda^\tau.$$

In turn, the expected number of new mutations in a given generation is well approximated by
$$E(Y_\tau(\tau)) = 2Nu.$$

It follows that the expected deleterious allele frequency is
$$E(q) = \frac{1}{2N} E(\sum_{\tau=1}^{\infty} Y_\tau(0)) = \frac{1}{2N} \sum_{\tau=1}^{\infty} E(Y_\tau(\tau))\lambda^\tau = \frac{u}{hs},$$

and thus the expected contribution to load is $2u$ - well-known results for mutation-selection balance.

Next, we consider a changing population size. We denote by $N(t)$ the population size $t$ generations in the past and by $a(t) = \frac{N(t-1)}{N(t)}$ the proportional change in one generation. Now the expected number of new mutations introduced at a given time is proportional to the population size
$$E(Y_\tau(\tau)) = 2N(\tau)u,$$

but the fraction of new mutations in the population remains constant ($u$). Similarly, the expected number of alleles in the next generation is affected by changes in population size
$$E(Y_\tau(t-1)) = \lambda a(t) E(Y_\tau(t)),$$

but their fraction is not, because their increase in number is precisely offset by the increase in population size
$$E(\frac{Y_\tau(t-1)}{2N(t-1)}) = \lambda a(t) \frac{N(t)}{N(t-1)} E(\frac{Y_\tau(t)}{2N(t)}) = \lambda E(\frac{Y_\tau(t)}{2N(t)}).$$

It follows that the proportional contribution of alleles to the present is the same as that in a constant population size:
$$E(\frac{Y_\tau(0)}{2N(0)}) = u\lambda^\tau,$$

leaving the deleterious allele frequency and the load at the present unchanged (at $\frac{u}{hs}$ and $2u$). In other words, the expected frequency of deleterious alleles and therefore the load follow the same deterministic dynamic as they do in a population of constant size, because when the population size changes, the increase (decrease) in the copy number is precisely offset by the increase (decrease) in population size.

We note that incorporating reverse mutation and migration will not change this conclusion. Reverse mutation would reduce $\lambda$, while introducing migration would be similar to both decreasing $\lambda$ (due to migration of deleterious alleles out of the population) and increasing the mutational input (due to migration of deleterious mutations into the population).

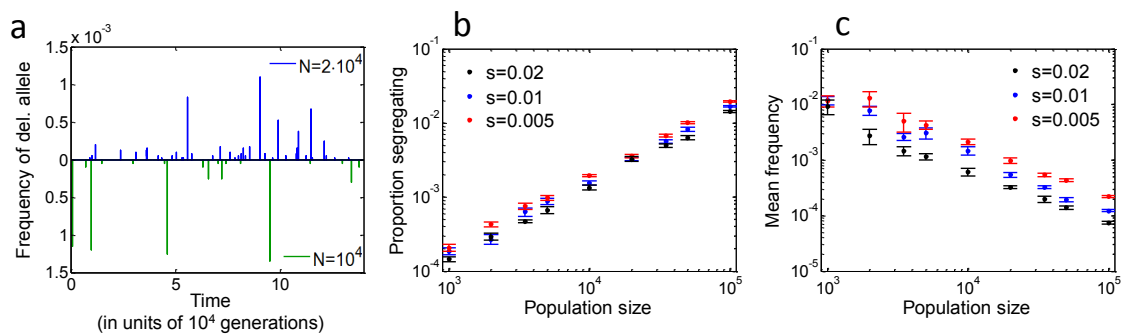

Figure 13: The equilibrium properties of segregating sites in the quasi-dominant case. In all graphs, $h = 0.5$ and $u = 10^{-8}$. A) Frequency of deleterious alleles as a function of time in simulations with two population mutation rates, corresponding to $N = 10^4$ and $2 \cdot 10^4$. In both cases, $s = 0.01$. B) The expected proportion of segregating sites as a function of population size. C) The expected frequency of deleterious alleles at segregating sites as a function of population size.

Our results clarify how the expected deleterious allele frequency and proportion of segregating sites at equilibrium depend on population size. When the population mutation rate is sufficiently low, a site switches intermittently between having no deleterious alleles and having a single mutation (by origin) in the population (Figure 13A). Under these conditions, in a larger population size, the mutational input is larger and thus the proportion of time that a site segregating increases (Figure 13B). Because the trajectory of a mutation in terms of numbers of copies does not depend on the population size, the frequency of the mutation is proportional to $1/N$, so the expected frequency of deleterious alleles at segregating sites scales with $1/N$ (Figure 13C). In turn, when the population mutation rate is sufficiently high, deleterious alleles are almost always present and often have several mutational origins. Under these conditions, the proportion of segregating sites approaches 1 (Figure 13B). Given that the expected frequency at segregating sites is $x = \frac{q}{S_{2N}}$, it follows that the allele frequency asymptotes to $q = \frac{u}{hs}$ (Figure 13C). In turn, the variance in allele frequency decreases with population size and asymptotes to 0 in the infinite population size limit.

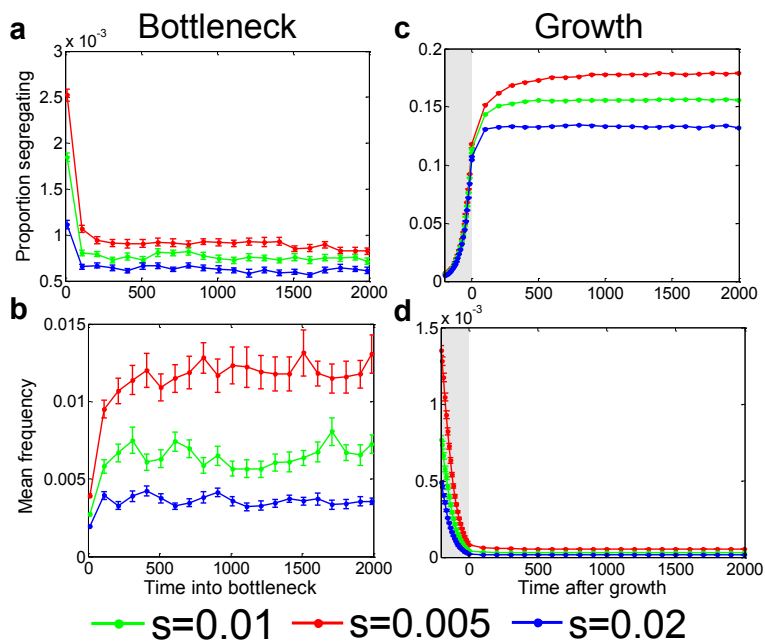

Figure 14: The properties of segregating sites as a function of time for the quasi-dominant case. In all graphs, $h = 0.5$. The proportion of segregating sites after (A) the reduction in population size in the bottleneck model and (C) the onset of growth. The expected frequency of deleterious alleles at segregating sites after (B) the reduction in population size in the bottleneck model and (D) after the onset of growth. The shaded region is the period of growth in the Tennessen model.

After a change in population size, a new equilibrium is attained much more rapidly in the strong selection regime because of the rapid turnover of deleterious alleles (see Figure 14). However, load is unaffected.

Thinking in terms of the branching process help us to evaluate previous conjectures about the possible effects of human growth on deleterious alleles. For example, Keinan and Clark [13] suggest that "Some degree of genetic risk for complex disease may be due to this recent rapid expansion of rare variants in the human population". It is indeed the case that the expected copy number of deleterious alleles should be greater under exponential growth; specifically, for a population growing at a geometric rate $\gamma$ per generation, the copy number will change at a geometric rate of $\lambda + \gamma$ per generation, which will result in an increase if $\lambda + \gamma > 1$. Moreover, population growth increases the sojourn time of a deleterious mutation and, when $\lambda + \gamma > 1$, there is a finite probability it would never go extinct [14]. Importantly, however, the expected *frequency* of quasi-dominant deleterious alleles remains constant, so human population growth has no effect on load.

**The recessive case**

In this case, the load at equilibrium is again insensitive to population size, but the underlying reasons are quite different than in the quasi-dominant case. In the recessive model, a deleterious allele behaves neutrally while at low frequencies. As a result, its sojourn time (i.e., the expected time that it spends at frequency x) is well approximated by that of a neutral allele (Figure 15B). When the frequency $x$ reaches $2Nsx^2 \approx 1$, selection on homozygotes for the deleterious alleles kicks in, and the allele should spend little time above this frequency. In the low mutation rate (LMR) approximation, we can therefore approximate the sojourn time of a recessive deleterious allele as

$$\tau(x) \approx \begin{cases} \frac{2(2N-1)}{1-x} & \text{if } 0 \leq x \leq \frac{1}{2N} \\ \frac{2}{x} & \text{if } \frac{1}{2N} \leq x < \frac{1}{\sqrt{2Ns}} \\ 0 & \text{if } \frac{1}{\sqrt{2Ns}} \leq x < 1 \end{cases},$$

where the expressions for $x < 1/\sqrt{2Ns}$ are the sojourn times (in generations) for a neutral allele (Fig 15B). In this approximation, the expected contribution of a deleterious mutation to load is then

$$s \int_0^1 x^2 \tau(x) \mathrm{d}x \approx s \int_0^{\frac{1}{\sqrt{2Ns}}} x^2 \frac{2}{x} \mathrm{d}x = \frac{1}{2N},$$

and, given that the expected input of new mutations per generation is $2Nu$, the overall expected load is

$$l(0, s) \approx 2Nu \frac{1}{2N} = u.$$

In other words, (in the low mutation limit) for a given population size $N$, a recessive allele behaves neutrally up to a frequency of $N^{-\frac{1}{2}}$, resulting in an expected contribution to load that is proportional to $N^{-1}$. In turn, the mutational input is proportional to $N$, so they exactly offset.

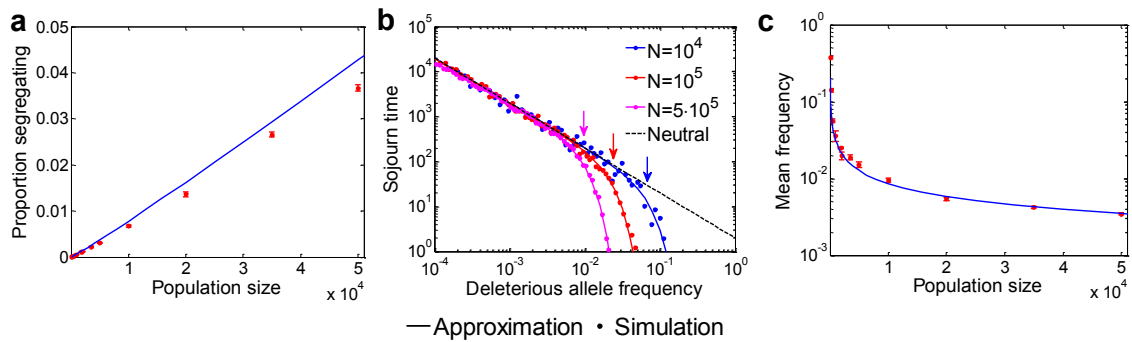

Figure 15: The properties of segregating sites at equilibrium in the recessive case, as a function of population size. The selection coefficient is $s = 0.01$. A) The proportion of segregating sites. B) The sojourn time of deleterious alleles for different population sizes. The threshold frequency of $\frac{1}{\sqrt{2Ns}}$ for each population size is marked by an arrow with the corresponding color. C) The average frequency of deleterious alleles.

This back of the envelope approximation also provides an intuitive explanation for the way in which the properties of segregating sites at equilibrium depends on population size (Fig 15). First, we consider the proportion of segregating sites (Fig 15A). When the population size is sufficiently small for the LMR approximation to apply, the proportion of segregating sites can be approximated by the ratio of the sojourn time of a single mutant through the population to the time between appearances of mutations, namely:

$$S_{2N} \approx \frac{\int_0^1 \tau(x)\mathrm{d}x}{\frac{1}{2Nu}} \approx 2Nu(ln(2N/s) + 2).$$

In a larger population size and hence with a larger mutational input, mutations of different origin will overlap, resulting in a slower increase in the proportion of segregating sites with population size. When the mutational input becomes sufficiently large, this proportion asymptotes to 1. Next, we consider the frequency of deleterious alleles. In the LMR approximation, the frequency spectrum of segregating sites can be approximated using the neutral sojourn times up to the threshold frequency $\frac{1}{\sqrt{2Ns}}$ (Fig 15B), yielding an average frequency of $E(x) \approx \frac{1+\frac{2}{\sqrt{2Ns}}}{2+ln\frac{2N}{s}}$. As the population size increases, such that mutations of different origins overlap, the decrease in average frequency becomes slower and asymptotes to $E(x) = E(q) = \sqrt{u/s}$ (Fig 15C). Lastly, the turnover time of segregating sites for a given population size $N$ is on the order of $2\sqrt{\frac{2N}{s}}$. As it was for other regimes; this is the time scale for the process of equilibration following a change in population size.

We now consider the implications for the bottleneck and growth models. In the bottleneck model, after the reduction in population size, there is an increase in load followed by a decrease back to the equilibrium level (Figure 16A). The transient increase in load (blue arrow in Figure 16A) is dominated by the contribution of mutations that segregated before the decrease in population size. The proportion of sites that segregated before was greater and their frequencies lower than after the population size reduction, and while these segregating mutations are gradually absorbed, some of them will drift to higher frequencies, generating a transient surge in load (Figure 16B). In turn, the newly introduced mutations have yet to reach equilibrium frequencies and, given that the contribution of the lower frequencies to load is much smaller, they contribute negligibly. In the Tennessen et al. model, the time that elapsed since the bottleneck is longer and the segregating sites are therefore closer to the new equilibrium (green arrow in Figure 16A). Correspondingly, the relative contribution of new mutations is greater and their frequency distribution is closer to equilibrium with the new population size, and yet some contribution from the older mutations remains (Figure 16C). These considerations also explain why load exceeds above equilibrium levels in the strong selection regime in Figure 6.

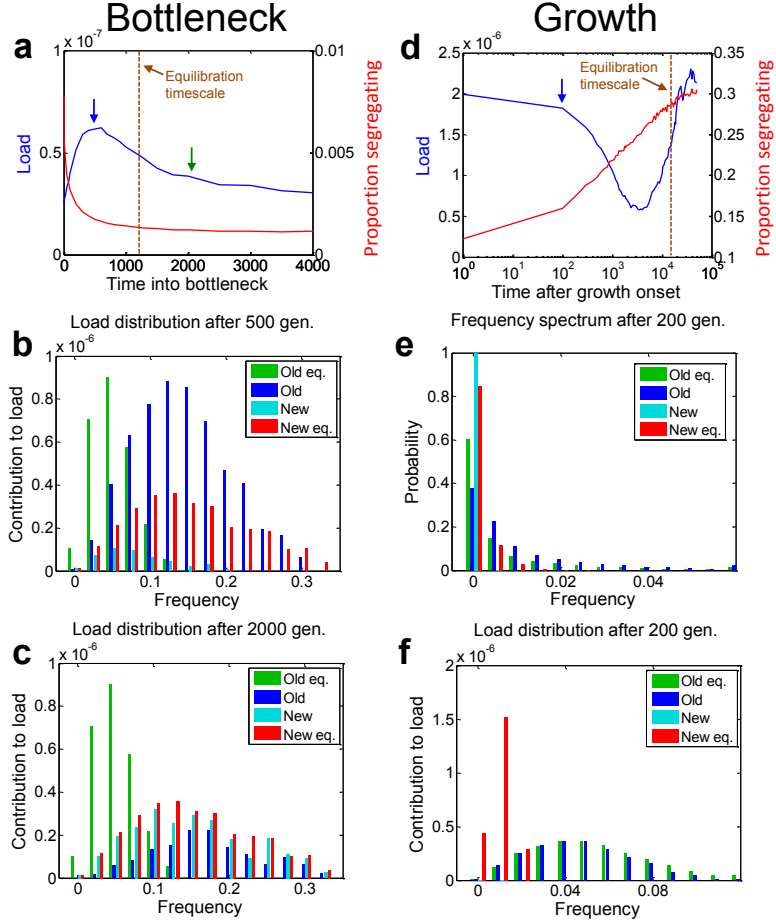

Figure 16: Load as a function of time in the recessive case. The selection coefficient is $s = 0.01$. A) The load and proportion of segregating sites as a function of time after the reduction in population size. B) The contribution to load of old and new mutations as a function of frequency, at the time of peak load (500 generations after the reduction in population size, indicated by a blue arrow in A). C) Same as B but for the time since the Out-of-Africa bottleneck, i.e., 50Kya (indicated by a green arrow in A). D) The load and proportion of segregating sites as a function of time after the onset of growth. E) The allele frequency distribution of old and new mutations at the end of the growth period (200 generations after onset, indicated by an arrow in D). F) The contribution to load of old and new mutations as a function of frequency at the end of the growth period.

In the growth scenario, we see the opposite transient effect: the load is reduced before recovering to its equilibrium level (Figure 16D). After the growth period, the number of segregating sites is greatly increased, but the new mutations have had little time to drift to higher frequency. As a result, new mutations segregate at very low frequencies and contribute negligibly to load (Figure 16E and F). In turn, mutations that segregated before growth have decreased in frequency due to the increased efficacy of purifying selection, and so their contribution to load declines substantially (Figure 16E and F). The result is a transient reduction in load (seen in Figure 6 as well as in Figure 16D).

## 2.4 Models with dominance coefficients other than $0$ and $\frac{1}{2}$

Here we provide summaries of simulations with dominance coefficients other than 0 and 1/2 to illustrate that the same qualitative behaviors are observed. As shown in Figure 17, all of the observed qualitative behaviors are included in our previous analysis and summarized in Table 1, with one possible exception.

The exception is in the bottleneck model in cases with dominance coefficients $h > 1/2$, where the total load is reduced for lower selection coefficients in the weak selection regime. The reason for this reduction in load is analogous to that for the increase in load that we saw in the recessive case in the same selection regime. For dominance coefficients greater than half, the extinction of low frequency deleterious alleles that segregated before the reduction in population size decreases load more than the fixation of high frequency deleterious alleles increases it. The opposite is true for dominance coefficients smaller than half.

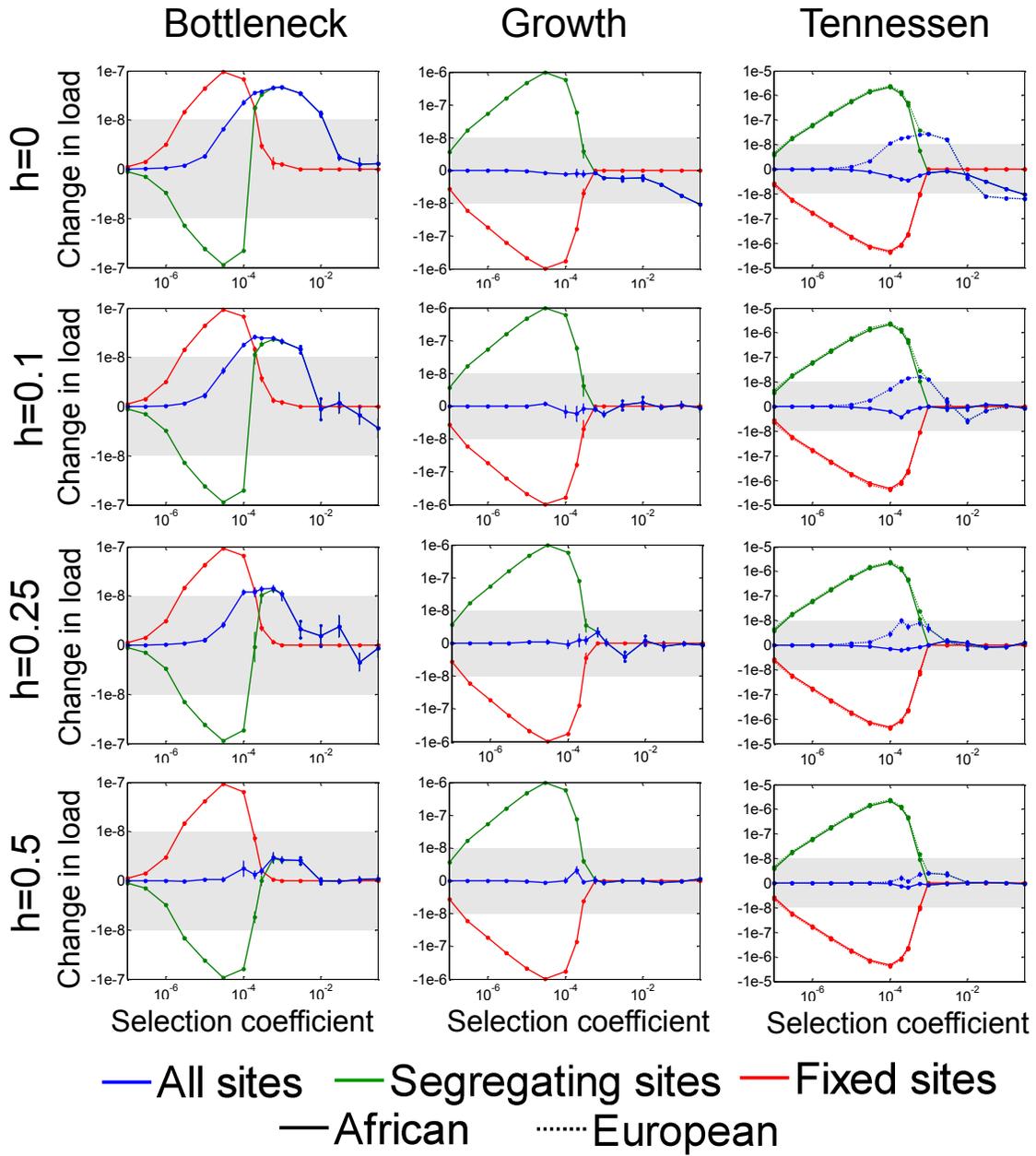

Figure 17: Continued on the next page.

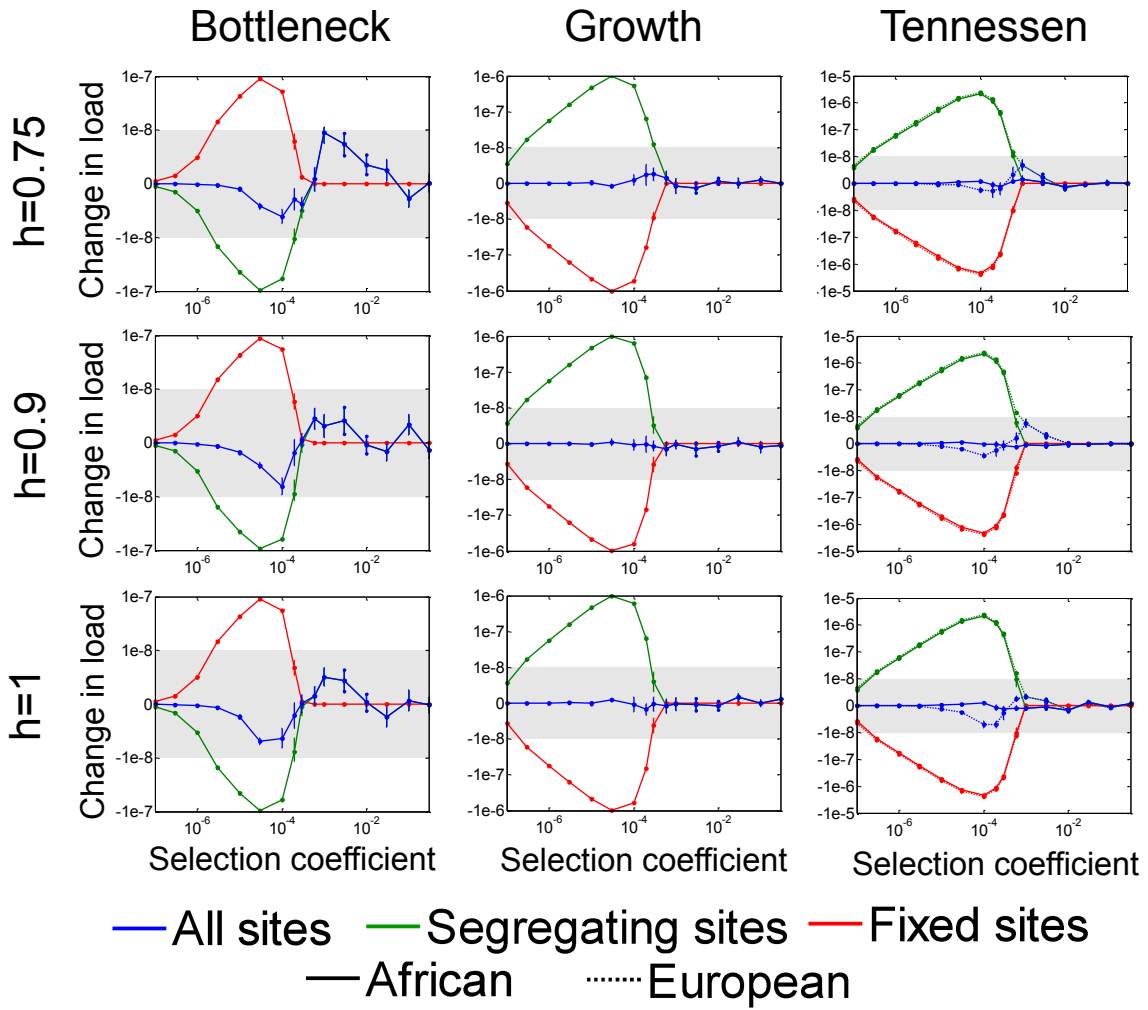

Figure 17: Changes in load under the three demographic models with different dominance coefficients. $h = 0$ and $1/2$ correspond to the results in Figure 6 and are provided for comparison.

# 3 Data analysis and interpretation

We used data from Fu et al. (2012) [4], who reported European-American (EA) and African-American (AA) allele frequency estimates and inferences of which allele is derived at each SNV (data available from `http://evs.gs.washington.edu/EVS/`). Variants for which allele frequencies are both 0 or both 1 in European- and African-Americans were dropped (presumably these are sites at which all sampled individuals differ from the genome reference sequence). We thank Josh Akey for assistance with accessing the data.

A natural test statistic for comparing the difference in load between two populations is to count the mean number of derived alleles per individual at SNVs segregating within the joint sample. Note that it is essential in these calculations to define SNVs using the joint sample, otherwise sites that are fixed for the derived allele in Population A but not in Population B would lead to the erroneous conclusion that that there are more derived alleles in B than in A. Another potential issue is that the number of derived alleles is not equivalent to the number of deleterious alleles, as a small fraction of weakly selected sites may actually be fixed for the deleterious allele. However, the qualitative predictions from our load models carry over to the derived allele measure. As shown in Figure 18, we predict that at semidominant sites there should be essentially no difference in mean frequency between AAs and EAs, regardless of selection coefficient. At recessive sites there may be a small increase in mean frequency in AAs at moderate and strongly selected sites. The fact that we do not observe any significant difference in allele frequencies at "probably damaging" sites argues that the majority of these sites are at least partially dominant.

Mean derived allele frequencies and their standard errors were calculated for both African Americans and European Americans at autosomal noncoding, synonymous, and nonsynonymous sites, as well as autosomal nonsynonymous variants belonging to the different functional categories. The ANNOVAR suite of scripts [19] was used to obtain functional predictions for each SNP from each of four prediction methods:

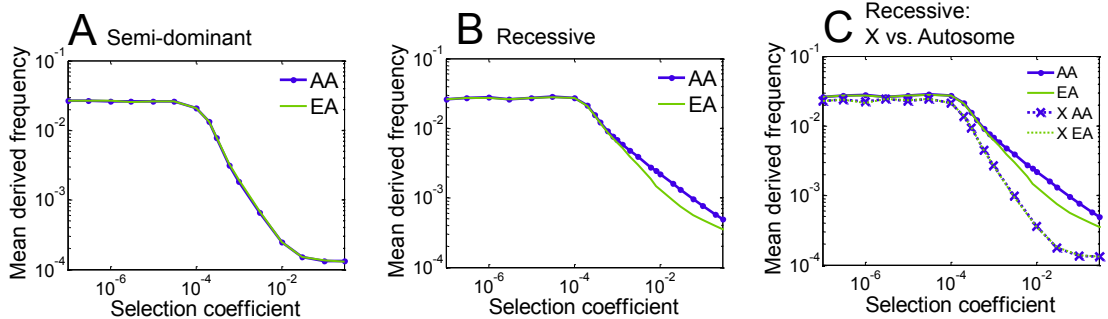

Figure 18: Mean derived frequencies predicted as a function of selection coefficient, for the AA and EA demographies. Notice that in (**A**) we predict that for semi-dominant sites AAs and EAs should have essentially identical mean derived frequencies for all levels of selection. In (**B**) we predict a small increase in mean frequencies for AAs at recessive sites with moderate-strong selection. (**C**) provides X vs autosome comparisons under the recessive model; note that recessive alleles on the X experience selection as dominant alleles in males.

PolyPhen2 [20], SIFT [21], LRT [23] and MutationTaster [22]. Default program settings were used in each case. The functional designations for each program are as follows: PolyPhen2: D (Probably Damaging), P (Possibly Damaging), B (Benign). SIFT: D (Damaging), T (Tolerant), LRT: D (Deleterious), N (Neutral) and U (Unknown). MutationTaster: A (Disease Causing Automatic), D (Disease Causing), P (Polymorphism Automatic) and N (Polymorphism). Coding versus non-coding and synonymous versus non-synonymous designations were also determined using ANNOVAR.

The haploid sample sizes in Fu et al were EA Autosomal: 8596, EA X: 6717, AA Autosomal: 4434, AA X: 3852. Our primary analysis in the main paper (reported in Figure 3) uses the full sample sizes with the autosomal data. For the purpose of Table 2 we wished to be able to compare means on the X and autosomes. Since mean allele frequencies of segregating sites are affected by total sample size, we implemented the following subsampling strategy to facilitate direct comparisons between X and autosomes. First, we converted the reported allele frequencies for each site back into allele counts (i.e., multiplying each reported frequency by the relevant haploid sample size). Next, we randomly subsampled the autosomal EA and AA variants and the X chromosome EA variant allele frequencies down to a sample size of 3852 chromosomes each, in order to match the haploid sample size for the African-American X chromosome. Subsampling was done without replacement, using the hypergeometric sampling function in R. After sub-sampling, variants whose allele frequencies were both either 0 or 1 were once again dropped. Two-sided t-tests were used to test for allele frequency differences between groups.

We observed that a strong reference bias exists at sites for which the genome reference sequence carries the derived reference allele. This bias has also been observed by David Reich and Shamil Sunyaev (personal communication). All four functional prediction programs designate a very high proportion of these sites as being likely nonfunctional or benign. When we condition on the overall population frequency at these sites, we find that a given site is much more likely to be classified as a probably damaging site if the reference genome carries the ancestral allele than if it carries the derived allele (Figure 19). To deal with this bias, we treated the functional designations at sites where the reference allele is derived as unreliable. As an alternative, we binned all SNVs into a series of allele frequency bins (i.e., the bins shown in Figure 19). We assumed that when we condition on the population allele frequency in a very large sample (i.e., the Fu et al sample) that the identity of the genome reference allele carries essentially no further information about the likely functional properties of a variant. Thus, within a bin, the fraction of derived-reference SNVs that fall into each functional category can be predicted from the fraction of ancestral-reference SNVs in that functional category. Thus for example, if 20% of the ancestral-reference SNVs in a given bin have functional category X, then we assume that each of the derived-reference SNVs in that bin has a 20% probability of also being in functional category X. The mean frequency of all SNVs in category X is estimated by summing across all ancestral-reference SNVs in category X plus a sum of contributions from all derived-reference SNVs, weighted by the estimated probabilities that each is in X. As shown in Table 3, the bias correction makes a substantial difference to the data analysis. Prior to applying the bias correction, the mean frequency in AAs is substantially higher than in EAs, but the bias correction makes the two frequencies virtually identical as predicted for models with dominance.

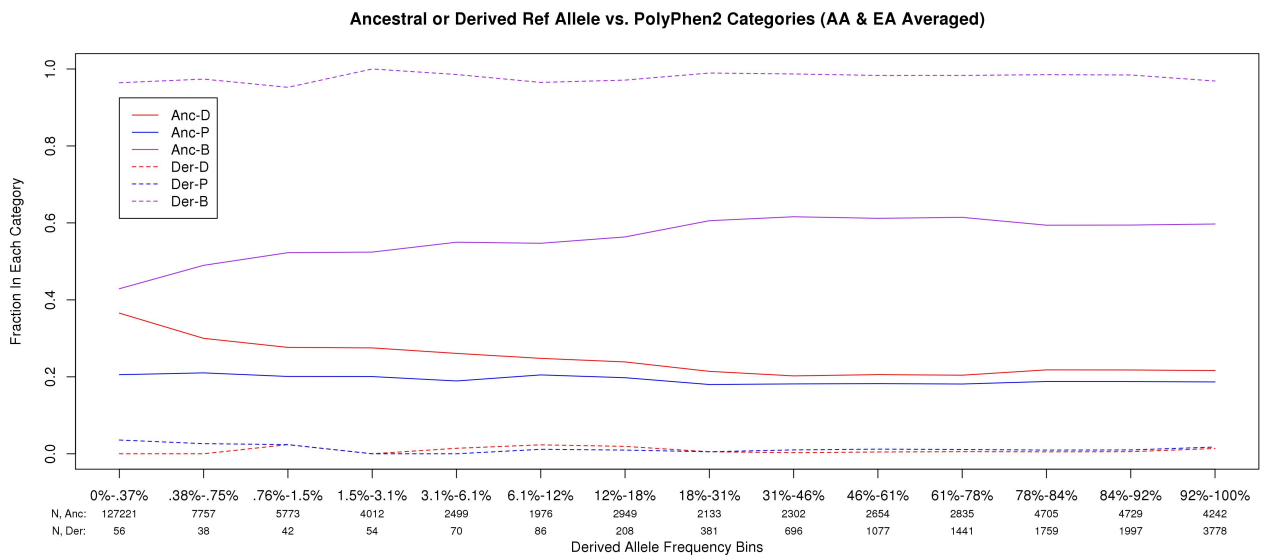

Figure 19: Illustration of the reference bias present in PolyPhen 2 [20]. The other functional prediction methods that we considered have a similar bias. The x-axis shows the mean population frequency of nonsynonymous SNVs in the Fu et al data (the left-most bins cover very narrow intervals of frequencies since most of the data are present in these bins). The y-axis plots the fraction of SNVs in each bin that are classified into each of the three PolyPhen categories: **B**enign, **P**ossibly damaging, Probably **D**amaging; and shown separately according to whether the genome reference sequence carries the ancestral or the derived allele. Notice that when the reference carries the ancestral allele, an SNV is classified as Damaging with a probability that ranges from nearly 40% at low frequencies to ≈20% at high frequencies (solid red line). In contrast, for SNVs where the reference carries the derived allele, the fraction of Damaging alleles is near 0% at all frequencies (dotted red line).

| Method | Chr. | Category | # SNVs | $AA_{Mean}$ | $AA_{SE}$ | $EA_{Mean}$ | $EA_{SE}$ | t-score |
|---|---|---|---|---|---|---|---|---|
| Non-coding | Aut | — | 300209 | 0.034 | 0.00026 | 0.034 | 0.00028 | 0.44 |
| Non-coding | X | — | 8355 | 0.030 | 0.0015 | 0.028 | 0.0016 | 1.1 |
| Synonymous | Aut | — | 220391 | 0.033 | 0.00030 | 0.033 | 0.00032 | 0.87 |
| Synonymous | X | — | 7001 | 0.028 | 0.0016 | 0.029 | 0.0018 | -0.10 |
| Non-synonymous | Aut | — | 351265 | 0.014 | 0.00015 | 0.014 | 0.00016 | 0.40 |
| Non-synonymous | X | — | 10293 | 0.012 | 0.00086 | 0.012 | 0.00095 | 0.076 |
| PolyPhen2 | Aut | D | 121280 | 0.0078 | 0.00011 | 0.0076 | 0.00012 | 1.2 |
| PolyPhen2 | Aut | P | 65400 | 0.012 | 0.00018 | 0.012 | 0.00020 | 0.52 |
| PolyPhen2 | Aut | B | 132047 | 0.019 | 0.00024 | 0.019 | 0.00026 | 0.55 |
| PolyPhen2 | X | D | 3205 | 0.0072 | 0.00065 | 0.0079 | 0.00078 | -0.99 |
| PolyPhen2 | X | P | 1957 | 0.013 | 0.0012 | 0.012 | 0.0012 | 0.98 |
| PolyPhen2 | X | B | 3948 | 0.014 | 0.0011 | 0.014 | 0.0012 | 0.044 |
| Sift | Aut | D | 145986 | 0.0095 | 0.00012 | 0.0093 | 0.00013 | 1.6 |
| Sift | Aut | T | 180091 | 0.018 | 0.00021 | 0.018 | 0.00022 | -0.13 |
| Sift | X | D | 4251 | 0.0099 | 0.00076 | 0.0096 | 0.00082 | 0.34 |
| Sift | X | T | 5517 | 0.017 | 0.0013 | 0.017 | 0.0015 | -0.29 |
| LRT | Aut | D | 146701 | 0.0060 | 8.5e-05 | 0.0060 | 9.5e-05 | -0.11 |
| LRT | Aut | N | 160179 | 0.020 | 0.00024 | 0.020 | 0.00026 | 0.20 |
| LRT | Aut | U | 13845 | 0.0066 | 0.00036 | 0.006 | 0.00039 | 2.6 |
| LRT | X | D | 3270 | 0.0038 | 0.00037 | 0.0034 | 0.00034 | 0.93 |
| LRT | X | N | 4548 | 0.017 | 0.0014 | 0.017 | 0.0016 | -0.37 |
| LRT | X | U | 886 | 0.0052 | 0.0013 | 0.0046 | 0.0015 | 0.40 |
| MutationTaster | Aut | D | 155138 | 0.0022 | 2.9e-05 | 0.0017 | 3.0e-05 | 18 |
| MutationTaster | Aut | A | 5089 | 0.00089 | 9.5e-05 | 0.00056 | 4.8e-05 | 4.3 |
| MutationTaster | Aut | N | 161169 | 0.0062 | 6.8e-05 | 0.0047 | 6.7e-05 | 21 |
| MutationTaster | Aut | P | 9040 | 0.36 | 0.0047 | 0.39 | 0.0051 | -6.5 |
| MutationTaster | X | D | 3860 | 0.021 | 0.0021 | 0.023 | 0.0023 | -1.2 |
| MutationTaster | X | A | 76 | 0.0010 | 0.00058 | 0.00039 | 0.00017 | 1.5 |
| MutationTaster | X | N | 5566 | 0.0030 | 0.00026 | 0.0013 | 0.00022 | 7.0 |
| MutationTaster | X | P | 131 | 0.16 | 0.028 | 0.16 | 0.029 | 0.28 |

Table 2: Comparison of mean frequencies in AAs and EAs at different classes of sites, classified according to whether the sites are on the autosomes or X, and using a variety of different functional classifications. For this table, the data were subsampled down to 3852 chromosomes for AAs and EAs each, to enable X vs autosome comparisons. Note that the mean frequencies in each row are not significantly different ($|t-score| < 2$, with the sole exception of the functional classifications from MutationTaster (which are highly significant). The unusual results for MutationTaster likely arise because MutationTaster uses previously estimated population frequencies in its classification, thus introducing biases for population genetic analysis.

| Method | Chr. | Category | Without bias correction | | | | With bias correction | | | |
|---|---|---|---|---|---|---|---|---|---|---|
| | | | $AA_{Mean}$ | $AA_{SE}$ | $EA_{Mean}$ | $EA_{SE}$ | $AA_{Mean}$ | $AA_{SE}$ | $EA_{Mean}$ | $EA_{SE}$ |
| Non-synonymous | Aut | — | 0.014 | 0.00015 | 0.014 | 0.000162 | 0.014 | 0.00015 | 0.014 | 0.00016 |
| PolyPhen2 | Aut | D | 0.0038 | 9.3E-05 | 0.0033 | 1.0E-04 | 0.0078 | 0.00011 | 0.0076 | 0.00012 |
| PolyPhen2 | Aut | P | 0.0060 | 0.00017 | 0.0053 | 0.00019 | 0.012 | 0.00018 | 0.012 | 0.00020 |
| PolyPhen2 | Aut | B | 0.026 | 0.00035 | 0.026 | 0.00037 | 0.019 | 0.00024 | 0.019 | 0.00026 |
| Sift | Aut | D | 0.0061 | 0.00013 | 0.0055 | 0.00014 | 0.0095 | 0.00012 | 0.0093 | 0.00013 |
| Sift | Aut | T | 0.020 | 0.00026 | 0.021 | 0.00028 | 0.018 | 0.00021 | 0.018 | 0.00022 |
| LRT | Aut | D | 0.0028 | 6.4E-05 | 0.0025 | 7.4E-05 | 0.0060 | 8.5e-05 | 0.0060 | 9.5e-05 |
| LRT | Aut | N | 0.023 | 0.00029 | 0.023 | 0.00031 | 0.020 | 0.00024 | 0.020 | 0.00026 |
| LRT | Aut | U | 0.0081 | 0.00048 | 0.0071 | 5.0E-04 | 0.0066 | 0.00036 | 0.006 | 0.00039 |
| MutationTaster | Aut | D | 0.0017 | 4.3E-05 | 0.0011 | 4.3E-05 | 0.0022 | 2.9e-05 | 0.0017 | 3.0e-05 |
| MutationTaster | Aut | A | 0.0013 | 0.00034 | 0.00099 | 0.00032 | 0.00089 | 9.5e-05 | 0.00056 | 4.8e-05 |
| MutationTaster | Aut | N | 0.013 | 0.00024 | 0.012 | 0.00025 | 0.0062 | 6.8e-05 | 0.0047 | 6.7e-05 |
| MutationTaster | Aut | P | 0.26 | 0.0027 | 0.30 | 0.0032 | 0.36 | 0.0047 | 0.39 | 0.0051 |

Table 3: Comparison of estimated mean frequencies in samples of 3852 chromosomes, with and without bias correction of the functional annotations. Recall that we observed that all four functional prediction methods typically have low probabilities of assigned 'damaging' status to SNVs where the genome reference carries the derived allele. Notice that prior to applying the bias correction, AAs tend to have higher allele frequencies at putatively damaging sites, as reported by Tennessen et al. This is likely because most of the reference genome is of non-African origin. After applying our bias correction, we observe that AAs and EAs have essentially identical allele frequencies in all functional categories (except for MutationTaster, likely for reasons discussed above).

# 4 The effects of demography on the genetic architecture of disease risk

A great deal of interest focuses on understanding how recent demographic history has affected the genetic architecture of disease and specifically whether the recent explosive growth has increased the contribution of rare variants to disease risk [15, 13, 16, 2]. Here, we use the theory that we developed to elucidate some of these effects.

## 4.1 A model relating allele frequencies to disease susceptibility

We first consider the relationship between selection on individual loci and disease risk. The few models for this relationship differ sharply in their assumptions. At one extreme, Pritchard [17] assumed that variants that increase disease susceptibility tend to be deleterious, but that otherwise there is no relationship between the strength of selection acting on these loci and the extent to which they increase disease susceptibility. In turn, Eyre-Walker [18] assumed a correlation between the strength of selection at a locus and its contribution to disease susceptibility. All else being equal, a stronger relationship between the disease risk and fitness implies that the variants that contribute more to disease risk are under stronger selection and, as a result, tend to be younger and rarer. It also follows that their frequency distribution would be more susceptible to the effects of recent demographic events. Here we consider models for the two extremes: one in which the effect sizes are independent on the selection coefficients and the other where the effect sizes are proportional to the selection coefficients.

To model how genetic variation relates to disease risk, we consider the $L$ loci that contribute to disease risk and denote the genotype of individual $i$ at these loci by $\mathbf{G_i} =$ $(g_{i,1}, \ldots, g_{i,L})$. We assume that each of the loci is bi-allelic, with a normal ($N$) and susceptible ($S$) alleles, and therefore denote the genotype at locus $j$ ($j = 1, \ldots, L$) as $g_{i,j} = NN$, $NS$, or $SS$. We then assume that the probability of developing the disease (ignoring life-history details) takes the form

$$P(\mathbf{G}) = F(\sum_{j=1}^{L} \alpha_j(g_j)),$$

where $F$ is a monotonically increasing function with continuous derivatives that takes values between 0 and 1 and that

$$\alpha_j(g) = \begin{cases} 0 & \text{if } g = NN \\ h_j a_j & \text{if } g = NS \\ a_j & \text{if } g = SS \end{cases},$$

where $h_j$ and $a_j$ denote the dominance coefficient and effect size of the contribution to susceptibility at locus $j$. Finally, we assume that the effect of each locus is small, such that we can approximate the variance in susceptibility by

$$V(P(G)) \approx [F'(\sum_{j=1}^{L} E(\alpha_j(g_j)))]^2 \sum_{j=1}^{L} V(\alpha_j(g_j)), \quad (5)$$

where the variances are taken over the population and

$$V(\alpha(g); x, a, h) = a^2 x(1-x) \left[(2h-1)^2 x^2 + (1-4h^2)x + 2h^2\right].$$

The model where the effect sizes are independent on the selection coefficients (and similarly for dominance coefficients) follows directly. For simplicity we assume that the effect sizes and dominant coefficients are constant, as assuming a distribution yields similar results for all the quantities that we consider below. In this case we simply assume that the $a_j$'s and $h_j$'s are constant across loci.

Now, we assume that the disease itself is the agent of selection, in other words that the fitness cost results entirely from the probability of developing the disease. Denoting the fitness of affected individuals by $W_a$ and of unaffected by $W_u$, the relationship between fitness, $W$, and the probability of developing the disease then takes the form

$$W = PW_a + (1-P)W_u.$$

In turn, in our model, the relationship between genotype and fitness is

$$W(\mathbf{G_i}) = \prod_{j=1}^{L} w_{i,j} \approx \exp\left(-\sum_{j=1}^{L} \alpha_j(g_{i,j})\right),$$

where

$$\alpha_j(g) = \begin{cases} 0 & \text{if } g = NN \\ h_j s_j & \text{if } g = ND \\ s_j & \text{if } g = DD \end{cases},$$

and we assume that $s_j \ll 1$ and therefore use an exponential approximation. Equating our two expressions for fitness leads to the following model for the relationship between disease risk and genotype

$$P(\mathbf{G}) = \frac{W_u - W(\mathbf{G})}{W_u - W_a} = \frac{W_u}{W_u - W_a} - \frac{1}{W_u - W_a} exp\left(-\sum_{j=1}^{L} \alpha_j(g_j)\right).$$

It follows that under this model, the dominance coefficient and effect size for the contribution to disease risk equal those for fitness (justifying our use of the same notation for the $\alpha$s in both).

We now return to the contribution of individual loci to disease risk under this model. Assuming that each locus has a small contribution, i.e., that $\alpha_j(g) \ll 1$ (which follows from $s_j \ll 1$) for $j = 1, \ldots, L$, we can approximate the variance in disease risk by

$$V(P) \approx \exp(-2\sum_{j=1}^{L} E(\alpha_j(g_j))) \sum_{j=1}^{L} V(\alpha_j(g_j)). \tag{6}$$

In other words, the contribution of an individual locus to variation in disease risk is proportional to the variance in fitness at that locus. Here, we consider semi-dominant and recessive loci for which the variances are

$$V(x; s, \frac{1}{2}) = \frac{1}{2}s^2 x(1-x) \tag{7}$$

and

$$V(x; s, 0) = s^2 x^2(1-x^2), \tag{8}$$

correspondingly.

## 4.2 Demographic effects on the variance

Figure 20 depicts how different allele frequencies at semi-dominant and recessive loci contribute to the variance in disease risk under the Tennessen et al. [2] model (expanding on Figure 3 in the main text). Because we consider only one selection coefficient at a time, the relationship between effect sizes and selection coefficient has no effect here; however, we do assume that the dominance coefficient for fitness and for disease risk are the same. The graphs can also be interpreted as the proportional contribution of different allele frequencies to the variance in fitness among individuals. To elucidate the effects of recent demographic events, we also show results for the model with a constant population size (equivalent to the one for the African population before the onset of growth) and for a population that experienced the same instantaneous increase in population size as the ancestral African population in the Tennessen et al. model but then remained constant (from $\sim 7,000$ to $\sim 14,500$ around $6,000$ generations ago, cf. Figure 1A), which we refer to as the older growth model.

**Demographic effects in the semi-dominant case.** First, we consider the effectively neutral regime (Figure 20A). In the model with constant population size, the proportional contribution is uniform across frequencies, as expected [3]. In the model of older growth, there is an increased contribution of low and high frequency alleles to the variance (as diversity patterns did not have sufficient time to reach equilibrium yet). In the model for Africans, a similar pattern is observed, with a tiny increase in the contribution from rare alleles due to recent growth (amounting to 0.41% of variance in deleterious variants with frequency below 0.1% and 0.4% in variants above 99.9%). In the model for Europeans, the increase due to growth is also negligible (0.61% of variance in variants with frequency below 0.1% and 0.6% in variants above 99.9%). However, the bottleneck leads to an increased contribution of intermediate frequencies at the expense of moderately low and high frequency alleles (since low and high frequency alleles are quickly lost or fixed after the reduction in population size).

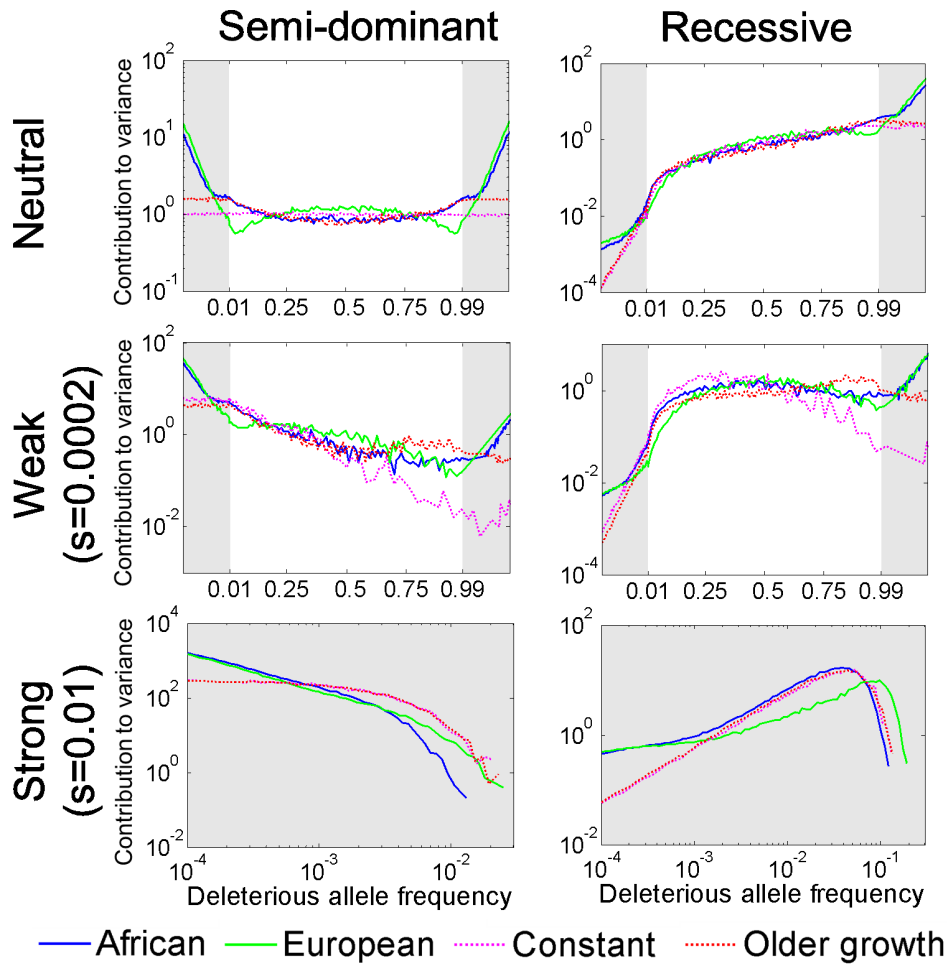

Figure 20: The proportional contribution of different allele frequencies to variance in disease risk, under the Tennessen et al. model for Africans and Europeans. Shaded regions correspond to a logarithmic scale on the x-axis, which is included to show the (minor) effects of recent growth.

In the weak selection regime (Figure 20B), selection leads to a shift towards lower frequencies and thus to an increased contribution to variance of lower frequency alleles. In turn, the effect of older growth is to increase the contribution of high frequencies: the reason being that before the increase in population size, a greater proportion of sites is fixed for the deleterious allele and at such sites, normal mutations lead to high frequency deleterious alleles. The recent growth in the model for Africans further causes a small increase in the contribution of rare alleles (amounting to 1.4% of variance in variants with frequency below 0.1% and 0.07% in variants above 99.9%). In the model for Europeans, this increase is also small (1.9% of variance in variants with frequency below 0.1% and 0.1% in variants above 99.9%), but the bottleneck again has a substantial effect, increasing the contribution of intermediate frequencies at the expense of lower and higher frequencies.

In the strong selection regime, because of the quick turnover of deleterious alleles, the older increase in population size and the bottleneck in Europeans are too far in the past to have had an effect on alleles that are currently segregating (Figure 20C). By the same token, in the Tennessen et al. model, alleles segregating at present are young and therefore the recent growth resulted in a decrease in their frequencies (cf. section 2.3), substantially increasing the contribution of rare alleles to variance (with $\sim 70\%$ of the variance contributed by alleles at frequency below 0.1%).

**Demographic effects in the recessive case.** In this case, recent growth has little effect in all selection regimes. The contribution of low frequency alleles to variance is much smaller because their effect on load or disease risk is manifested only in homozygotes (Figure 20D-F). As a result, the increase in the number of rare deleterious alleles caused by recent growth has a negligible effect on their contribution to the variance in disease risk under both the model for Europeans and Africans (amounting to $\sim 10^{-4}\%$ in the neutral regime, $\sim 5 \cdot 10^{-4}\%$ in the weakly selected and $\sim 0.01\%$ in the strongly selected regime, in variants with frequency below 0.1%). In turn, the increase in the number of high frequency alleles (due to normal mutants on a deleterious background) has a higher impact but it is still quite small (amounting to $\sim 1\%$ in the neutral regime and $\sim 0.2\%$ in the weakly selected regime that are due to variants with frequency above 99.9%).

In the weak and strong selection regimes, there is a peak in the contribution to variance at intermediate frequency (Figure 20E and F). Moving from low to intermediate frequencies, the contribution to the variance of a mutant allele increases (see Equation 8). This increase is halted, however, because at higher frequencies, selection on homozygotes for the deleterious allele kicks in, leading to few alleles at high frequencies. (Specifically, for a constant population size and given a low mutation rate, the frequency spectrum of deleterious alleles is well approximated by $C\frac{e^{-\alpha x^2}}{x}$, where $C$ is a normalizing constant [3], and thus the contribution to variance can be approximated by $De^{-\alpha x^2}x(1-x)^2$, where $D$ is a normalizing constant.) In the model for Africans (and for older growth), this peak is at higher frequencies in the weak selection regime (Figure 20E), because the older increase in population size led to relatively more high frequency alleles at present.

The bottleneck in the model for Europeans has a much more pronounced effect, causing a shift toward intermediate allele frequencies and a corresponding shift in the contribution to variance in all selection regimes (Figure 20D-F). As opposed to the semi-dominant case, this is also true for the strong selection regime, as recessive deleterious alleles can reach substantial allele frequencies.

**Summary.** Population growth increases the relative proportion of rare alleles and could therefore be expected to increase their relative contribution to the variance in disease risk. However, because rare alleles contribute less to the variance to begin with, this effect may be relatively small. Assessing the effects of growth on the genetic architecture of disease risk therefore requires quantification. Here, we have shown that, at least based on current estimates of recent growth, the effects on the variance in disease risk are expected to be negligible. The one exception is the case of strongly selected quasi-dominant alleles, which are young and therefore whose frequencies do reflect the recent population size expansion. Interestingly, in this case, while the architecture of disease risk is substantially affected by growth, the expected load (or disease prevalence) remains unchanged, i.e., the same load will be due to many more deleterious alleles that segregate at lower frequencies than had the population not grown.

In contrast to growth, the bottleneck in European populations should have increased the proportion of intermediate frequency deleterious alleles at the expense of low and high frequency ones (with the exception of strongly selected quasi-dominant alleles, because they are so young). In other words, in these populations, there will be only a small effect on load but a substantial effect on the architecture of disease, with a greater proportion of the variance in disease risk due to intermediate frequency alleles.

## 4.3 The contribution of rare alleles in a mixture model

In reality, we expect that the variants underlying a complex disease will have a variety of selection coefficients and effect sizes rather than a single one. Under a model with such a mixture, the expected contributions of different allele frequencies to the variance in disease risk can be derived as follows. For simplicity, assume that mutations are semi-dominant (so the dominance coefficient is dropped from the notation). At a site with selection coefficient $s$, the expected contribution to the variance from deleterious alleles below frequency $\omega$ is

$$V_\omega(s) = \frac{1}{2} C E(a^2|s) \int_0^\omega f(x;s)x(1-x)dx, \tag{9}$$

where $E(a^2|s)$ is the expectation of the effect size squared for sites with selection coefficient $s$, $f(x;s)$ is the probability of the deleterious allele being at frequency $x$ (here, we do not condition of the allele segregating) and the proportion coefficient $C$ is akin to the first term in Equation 5. The overall contribution to variance of a site is $V_1(s)$ and the fraction of that contribution coming from variants below frequency $\omega$ is $\Theta_\omega(s) \equiv \frac{V_\omega(s)}{V_1(s)}$. When all sites are considered jointly, denoting the input of mutations with selection coefficient $s$ by $\mu(s)$, the expected proportion of variance from deleterious alleles below frequency $\omega$ is then

$$\Theta_\omega = \frac{\int_s \mu(s) V_1(s) \Theta_\omega(s) ds}{\int_s \mu(s) V_1(s) ds}. \tag{10}$$

Examining the terms in Equation 10 suggests that the contribution of rare alleles depends strongly on the relationship between effect sizes and selection coefficients. Specifically, the proportional contribution of rare alleles $\Theta_{0.1\%}(s)$ becomes substantial only for strong selection coefficients (Figure 4D in the main text), as shown in section 4.2. The behavior of the overall contribution to variance $V_1(s)$, however, depends on the relationship between effect sizes and selection coefficients. If we assume that the effect sizes do not depend on the selection coefficients (or more precisely that $E(a^2|s)$ is constant) then $V_1(s)$ from weakly selected sites is much greater than from strongly selected sites (Figure 4E in the main text) and rare alleles will make an important contribution only if a very large fraction of the mutational input is at strongly selected sites. If we assume the other extreme in which the effect sizes are proportional to the selection coefficient (or more precisely that $E(a^2|s) \propto s^2$, as in the model in section 4.1) then $V_1(s)$ strongly increases with the $s$ (Figure 4E in the main text) and rare alleles would make an important contribution unless the fraction of the mutational input at strongly selected sites is very small. In reality, the outcome could be anywhere in between.

As an illustration, we consider a simple example, in which a fraction $\alpha$ of the sites have a weak selection coefficient $s_w = 0.0002$ and a fraction $1 - \alpha$ have a strong selection coefficient of $s_s = 0.01$. The contribution of rare alleles (below $\omega = 0.1\%$)

takes the form

$$\Theta_\omega(\alpha) = \frac{\alpha V_1(s_w)\Theta_\omega(s_w) + (1-\alpha)V_1(s_s)\Theta_\omega(s_s)}{\alpha V_1(s_w) + (1-\alpha)V_1(s_s)}. \tag{11}$$

With equal effect sizes, rare alleles make an important contribution only if a very large fraction of the mutational input is at the strongly selected sites, while the converse is true if the effect sizes are proportional to the selection coefficients (Figure 4F of the main text).